\documentclass{article}
\usepackage{graphicx} 
\usepackage{amsmath}
\usepackage{authblk}
\usepackage[a4paper, margin = 1.3in]{geometry}
\usepackage{float}
\usepackage{cite}
\usepackage{changepage}
\usepackage{float}
\usepackage[caption=false]{subfig}
\usepackage[font=scriptsize]{caption}
\usepackage[colorlinks=true, allcolors=blue]{hyperref}

\newtheorem{definition}{Definition}%

\def\BibTeX{{\rm B\kern-.05em{\sc i\kern-.025em b}\kern-.08em
    T\kern-.1667em\lower.7ex\hbox{E}\kern-.125emX}}
    
\providecommand{\keywords}[1]
{
  \small	
  \textbf{\textit{Keywords---}} #1
}

\title{Trends and Topics: Characterizing Echo Chambers' Topological Stability and In-group Attitudes}

\author[1,2]{Erica Cau}
\author[1,2]{Virginia Morini}
\author[2]{Giulio Rossetti}

\affil[1]{Computer Science department, University of Pisa, Italy }
\affil[ ]{\{erica.cau, virginia.morini\}@phd.unipi.it}
\affil[3]{KDD Laboratory, ISTI, National Research Council, Pisa, Italy}
\affil[ ]{{ \{virginia.morini, giulio.rossetti\}@isti.cnr.it}}

\date{}
\begin{document}

\maketitle
\begin{abstract}
Social Network sites are fertile ground for several polluting phenomena affecting online and offline spaces. 
Among these phenomena are included echo chambers, closed systems in which the opinions expressed by the people inside are exacerbated for the effect of the repetition, while opposite views are actively excluded. 
This paper offers a framework to explore, in a platform-independent manner, the topological changes of echo chambers through time while considering the content posted by users and the attitude conveyed in discussing specific controversial issues.

The proposed framework consists of four steps: (i) data collection and annotation of users' ideology regarding a controversial topic, (ii) construction of a dynamic network of interactions, (iii) echo chambers extraction and analysis of their dynamics, and (iv) topic extraction and valence analysis. 
The paper then enhances the formalization of the framework by conducting a case study on Reddit threads about sociopolitical issues (gun control, American politics, and minorities discrimination) during the first two years and a half of Donald Trump's presidency. 

The results unveil that users often stay inside echo chambers over time. 
Furthermore, the analyzed discussions focus on controversies related to right-wing parties and specific events in American and Canadian politics. 
The analysis of the attitude conveyed in the discussions shows a slight inclination toward a more negative or neutral attitude when discussing particularly sensitive issues, such as fascism, school shootings, or police violence.  
\end{abstract}


\section*{Introduction}
\label{sec:introduction}
The emergence of Online Social Network sites (OSNs) in the \textit{information age} ~\cite{Floridi2014} reshaped every aspect of life, spanning from those we show in the infosphere to how we communicate, with no longer any geographical and temporal constraints. 
Moreover, owing to the lack of these limitations, the Internet has made the exchange of opinions between users immediate, and the same has occurred with information that is shown to users instantaneously. 
Hence, the disclosure of new problems unknown in the past. 
An example is the \textit{information overload} to which users are exposed when accessing online spaces. 
The massive amount of conflicting information found online may lead users to experience a mental discomfort called \textit{cognitive dissonance} \cite{Festinger1962}.
Consequently, to avoid this discomfort, people are more prone to filter and choose only pieces of information confirming their beliefs and ideas, helped by recommendation systems introduced in OSNs.
\\
However, despite the importance of opinion heterogeneity for creating meaningful debates -- and consequently allowing the unfolding of the dialectic process of \textit{thesis, antithesis, and synthesis} -- at the same time, OSNs represent a perfect breeding ground for both human and algorithmic biases which may interfere with discussions and knowledge formation. 
The rise and, most of all, the exacerbation of these biases contribute to creating pollution in online spaces. 
Furthermore, this issue has grown in importance because of the loss of an evident boundary between the online and offline world, thus resulting in potentially harmful consequences that may easily overflow into the real world. 
Among the pollutants, polarization has raised several concerns due to the features offered by OSNs, as they tend to exacerbate the ideological positions of users by allowing for easier connections with people with the same interests and exposing them to content aligned with their thoughts. 
\\
This work focuses on one of the main consequences of polarization: the widely debated concept of \textit{echo chamber} (henceforth, EC). 
Although there is no agreement on a formal definition of echo chambers, their effects are noticeable. They are often argued to be involved in spreading misinformation, promoting pseudo-science narrations, and exacerbating political discourse. 
The discussions behind \textit{what} is an echo chamber picture toward echo chambers as a closed system where opinions and ideas are reinforced and exacerbated as the only truthful view of reality, for the effect of the repetition inside this environment and the active exclusion of other -- opposing -- beliefs. 
Existing works to date have analyzed the effects of ECs and assessed their presence in online spaces, but often from a modelistic perspective or through case studies that generally do not consider their temporal unfolding. 
\\
The aim of this work is threefold. 
First, it formalizes a platform-independent framework for investigating the dynamics of the relations that have often been ignored in the literature. 
Second, it defines a methodology to investigate the topics and the attitude of users inside and outside ECs after their identification. 
Third, it enriches the body of work on ECs dynamics by offering a case study on Reddit sociopolitical discussions during the first two years and a half of Donald Trump's presidency. 
Here, the ECs are detected, tracked through time, and analyzed by considering the content and investigating the emotional component of the discussions. Additionally, we compare ECs and less polarized debates to gain insights into the possible similarities and differences in their behavior over time.
\\
The proposed framework is an expansion of the previous framework for ECs detection described in \cite{Morini2021}, as it adopts the core of that work, that is, to develop a framework built on top of common features among OSNs while leveraging network science structures and algorithms to perform ECs detection. 
The framework we present is instead composed of four steps, where the first three are related to the topological and ideological identification over time of ECs, while the latter deals with the characterization of the contents discussed and the emotional characterization of users' discussions both inside and outside ECs. 
\\
The paper is organized as follows. 
In the Related works, we introduce and discuss the literature on ECs, focusing on their detection. 
Subsequently, the Framework section describes the proposed framework for tracking and analyzing ECs dynamics over time, emphasizing both the relations and topics of discussion. 
The framework is tested on OSN data extracted from Reddit in the Case Study section, and the results obtained are discussed. 
The Conclusion section summarizes the main findings of this project and discusses weaknesses and future developments. 

As the concept of ECs itself is widely discussed, there is also much debate on how these polarized systems create and develop.
This information is necessary to allow their recognition and, subsequently, their efficient mitigation to avoid potentially harmful outcomes. 
In the last decades, an ever-growing body of research has focused on quantifying the extent to which discussions are polarized \cite{Conover2021, Adamic2005, Guerra2021} and, consequently, are deemed to be a fertile ground for polluted phenomena. 
Usually, ECs primarily originate in online discussions about controversial topics in OSNs. Still, traces of ECs have also been found in forums \cite{Edwards2013}, blogs \cite{Gilbert2009}, and, generally, in those online spaces employing recommendation systems, such as e-commerce platforms \cite{Ge2020}.
\\
Traditional EC detection methods may follow two different families of approaches.
In the first case, the focus may be on the textual content shared by users (i.e., the \textit{echo} dimension), which corresponds to the debated opinion echoing among people with the same ideological alignment. 
Examples of this methodology include estimating users' views conveyed through words without considering users' relations. For example, in \cite{An2014}, the authors attempted to understand whether users were exposed to crosscutting content by investigating the news articles they shared on Facebook.
Another example can be found in \cite{Bakshy2015}, in which over 10 million U.S. Facebook users with public political leaning in their profile information were classified into two categories. 
In \cite{Caldern2019}, the authors define two methods to gain insight into the content of ECs: one to estimate the stance on a particular topic and another to determine the type of emotion and its intensity. 
The main issue of content-based approaches is that the data needs to be annotated, and this is often performed via unsupervised Natural Language Processing techniques, which does not ensure the correctness of the labels.
\\
The other family of methods to address this issue considers the network knit by users talking between themselves in OSNs discussions. 
In this scenario, the \textit{chamber} dimension is investigated, which opens up the reverberation of the opinion. 
The analysis consists of modeling the graph to inspect users' relations at various topological levels (i.e., \textit{retweet network}, \textit{comment network}), eventually with the possibility of enhancing the analysis through the mining of additional information from the text, such as in \cite{Garimella2018}, where Garimella \textit{et al.} estimated the users' political leaning and then proceeded with the construction of the interaction network, defining different roles for users in ECs formation.

The large number of network-based approaches could be further grouped based on the \textit{topological} scale of detected ECs: \textit{macro-scale}, \textit{meso-scale}, and \textit{micro-scale}. 
The first examines users' relations as a whole. It involves looking at the interaction networks on an aggregate level to identify two well-distinguished clusters of users with opposite leaning in the network. 
For example, in \cite{Kratzke2023}, the study extracted two communities using the HITS algorithm without leveraging any other information from external analysis, e.g., NLP. 
Another macro-scale study can be found in \cite{DeFrancisciMorales2021}, in which the authors reconstructed the interaction network and then analyzed the interactions between Donald Trump and Hillary Clinton supporters.
The meso-scale approach, instead, looks more in-depth at the topological division of nodes into clusters, usually by leveraging a community detection algorithm, to detect echo-chamber-like structures composed of nodes sharing identical ideological leaning. 
An example of a hybrid meso-scale and content-based approach was described in \cite{Villa2021}, where the authors explored the presence of ECs in tweets about COVID-19 by constructing the interaction network and by applying a community detection algorithm (METIS), which allowed to partition the network into two distinct communities. 
The communities were then evaluated according to different measures, both traditional community evaluation measures and controversy measures.

Ultimately, the last category consists of investigating the leaning of each user and comparing it to the one adopted by the members of the neighborhood, such as in \cite{Cinelli2021}, where the authors leverage homophily to assess the presence of ECs, moved by the idea that users surrounded by people with a similar leaning are consequently exposed to similar content(s), thus increasing the likelihood of ECs formation.

One issue regarding these studies is the methodology employed. They are often structured as data-driven case studies that assess the presence of ECs in controversial discussion topics without specifically focusing on analyzing the content or the emotional counterpart.
Another issue is that they usually rely on platform-specific features, thus making these methods difficult to reuse on other platforms to perform the same task. 
In addition, it is often neglected another invisible yet fundamental component of every complex system: time. 
This issue is usually approached through case studies revolving around a short timespan or from a merely modelistic perspective and is addressed as an \textit{opinion dynamics} task. 
Regrettably, such flattened representations, keeping together interactions potentially distant in time and disregarding their temporal ordering, describe a complex phenomenon in a simplistic manner, risking an overestimation of users’ sociality and failing to understand the real dynamics behind their appearance and evolution.

\subsection*{Dynamic community detection}\label{sec:relatedWorksDCD}
The approaches performing ECs detection at a meso-scale topology level usually rely on community detection (henceforth, CD) algorithms, which can detect homogeneous clusters of users sharing common features. 
Even if there is still no agreement on what a \textit{community} should be, several algorithms have been proposed to identify communities and, most importantly, to track their temporal unfolding. 
The attempt to have a glimpse into community dynamics adds another layer of complications because nodes and edges between users may undergo different events \cite{Palla2007}. 
For example, the disappearance of a node or a link usually leads to a topological variation that contributes to its lifecycle \cite{Cazabet2019}. 
While there are case studies employing CD algorithms in the context of polarized information systems detection \cite{Conover2021}, there is a scarcity of works employing dynamic community detection (DCD) to define frameworks or to present case studies about specific issues. 
Among the exceptions, there is the work by Kopacheva \textit{et al.} \cite{Kopacheva2022}, where the authors leverage DCD over a timespan ranging from 2012 to 2019 to analyze the evolution of users' communities on Twitter revolving around the refugee crisis in Sweden in 2015.

Because users inside an EC are involved in debates that bring out their opinions, it is also necessary to consider the ideology of each user when modeling the interaction network and when performing the DCD task. 
This can be accomplished by modeling the interaction network using a \textit{node-attributed graph}, where each node is associated with an attribute expressing users' leaning. 
Formally, $\mathcal{G} = (V, E, \Lambda)$, where \textit{V} is the set of vertices, \textit{E} is the set of edges, and $\Lambda$ is the set of \textit{m} attributes associated with vertex \textit{v} $\in V $ to represent attributes. 
In addition, for the extraction of communities in a network snapshot, we will leverage \textit{Labeled Community Detection}, a specific instance of CD that considers both topological criteria and label homophily inside each community when extracting the partitions.

\subsection*{Topic modeling and valence analysis}\label{sec:relatedWorksTopicModeling}
Moved by the intention to analyze ECs in more depth once they have been identified, in this section, we briefly present the state of the art of the two approaches employed to gain insights about the topics and users' attitudes when discussing online.

\textbf{\textit{Topic modeling.}  } The first approach, namely \textit{topic modeling}, is related to the extraction of the most relevant information, or topics, from a textual \textit{corpus}, i.e., from a collection of documents. 
The potential and usefulness of this task have been widely recognized and employed in the literature, even in fields of research not immediately related to linguistics, e.g., in bioinformatics \cite{Liu2016} or in computer vision \cite{Hospedales2011}. 
Many approaches and algorithms have been formalized and implemented to address this task. 
Among these approaches, one of the first and most well-known is Latent Dirichlet Allocation (LDA) \cite{Blei2003}, a probabilistic model that assumes that a statistical process generates each document. 
Therefore, each document has its own distribution of topics, and consequently, each topic would be characterized by the probability of its specific words. 
LDA, for its own nature, lacks semantic information and disregards relations between words and syntactic structures.
\\
In recent years, many evolutions of LDA have been proposed, such as LDA2vec \cite{Moody2016}, which incorporates Word2Vec \cite{Mikolov2013} into the LDA model. 
Topic modeling algorithms have also evolved to capture the dynamics of topics in documents over time, i.e., DTM \cite{Blei2006} and TTM \cite{Iwata2009}.
As an effect of the development of deep learning models, \textit{Transformers} have revolutionized the way documents are represented as vectors. 
In 2018, Bidirectional Encoder Representations from Transformers (BERT) \cite{Devlin2018} was released. 
Since then, considerably improved models have been released, such as RoBERTa \cite{Liu2019}, BART \cite{Lewis2020}, and DistilBERT \cite{Sanh2019}. 
These models are well known for being trained over a massive amount of data and for implementing the so-called \textit{attention} mechanism \cite{Vaswani2017}, which builds for each word a particular and contextual word embedding while accounting for both the words to its right and its left (thus, define BERT as a \textit{bidirectional} model). 
This revolution in language representation models has led to their massive application in several NLP tasks, e.g., text generation, classification tasks, and Named-Entities Recognition. 
\\ \ \\
\textbf{\textit{Valence analysis. }}
Valence has often been investigated as one factor defining the meaning of words and has been addressed as a useful measure in Natural Language Processing, psychology, and cognitive sciences. 
Valence (V) is often paired with two other meaning-related dimensions, i.e., Arousal (A) and Dominance (D). 
According to \cite{Russell2003}, different values in these measures might be employed to extract primary emotions. 
Typically, measures quantifying Valence, Arousal, and Dominance are extracted through manually annotated datasets, such as in the case of ANEW \cite{Bradley1999} and its extension by Warriner \textit{et al.} \cite{Warriner2013}. 
Another dataset is the VAD lexicon \cite{Mohammad2018}, which consists of around 20.000 English words manually annotated. 
Their ratings were aggregated by leveraging the Best-Worst scaling \cite{Louviere2015}.
The resulting values are in the range [0, 1], namely, the negativeness and positiveness of the concept indicated by the word.

\section*{Echo chambers diachronic analysis}\label{sec:Framework}
In this section, we formalize a platform-independent framework for tracking and analyzing the dynamics of ECs, while also addressing the content-related aspect of the issue. 
Note that the first steps of the pipeline consist of the platform-agnostic framework from Morini \textit{et al.} \cite{Morini2021}, to which we add two additional steps to handle the temporal analysis and the content characterization of ECs. 
The original framework that we enhance lies within approaches that investigate meso-scale topologies. This fits well with the leading theory about ECs, according to which they are close systems of users discussing mostly between themselves, with few social interactions with those outside.
Before defining the four-step framework, it is worth properly defining the model we use to represent online debates and, as a consequence, what we consider to be \emph{echo chambers}.

\begin{definition}[Interaction Graph]
\label{def:graph}
     Online debates are modeled as feature-rich networks $G=(V,E,A)$ describing a set $V$ of users interacting on a controversial topic $\eta$ (thus establishing dyadic edges $(u,v)\in E, u,v\in V$) each having an opinion $A(v)$ on $\eta$. 
\end{definition}

\begin{definition}[Echo Chamber]
    Given an interaction graph $G=(V,E,A)$, an echo chamber is a densely connected subset $E_i \subseteq V$, composed by users' sharing a same opinion $a_i\in A$ on $\eta$ that at the same time is loosely connected w.r.t. the rest of the network.
\end{definition}

\begin{figure*}
    \centering
    \includegraphics[scale = 0.4]{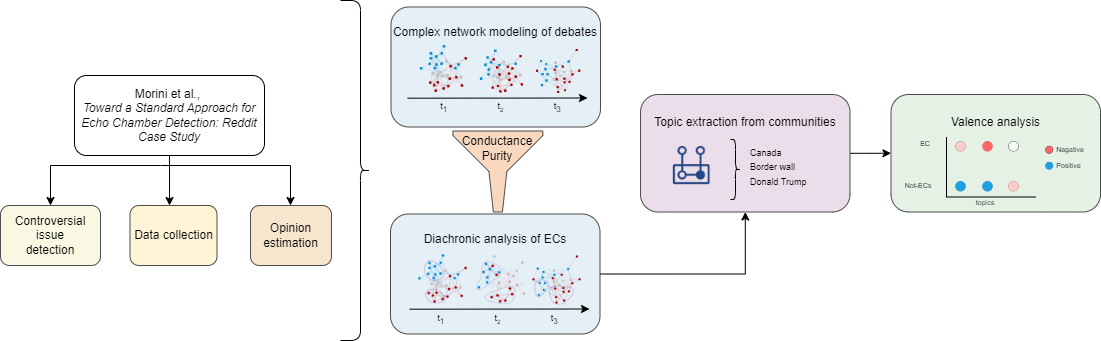}
    \caption{Graphical representation of the four steps of the framework.}
    \label{fig:framework}
\end{figure*}

Leveraging such modeling framework, our pipeline version collapses the four steps described in \cite{Morini2021} in two and proceeds by formalizing two new additional steps. 
EC detection and analysis is then structured as follows: (i) data collection and opinion estimation in the context of online polarized discussions, (ii) complex network modeling of online debates, (iii) users' groups' identification (by means of Dynamic Community Detection), and lifecycle analysis, and (iv) topic extraction and valence analysis.
A graphical representation of the proposed framework can be found in Figure ~\ref{fig:framework}.

\subsection*{Step 1: Data collection and annotation}
The starting point of the original pipeline is the identification of a controversial issue regarding a wide variety of topics, ranging from politics to social and environmental issues. 
This is necessary because users are more prone to assume an extreme ideology when discussing controversial topics, as they form the fertile ground for echo chamber formation. 
Moreover, users often use particular hashtags to easily identify controversies on the OSN, such as on Twitter, whereas on Reddit, users may join topic-based communities of discussion. 
Once the data have been obtained, it is necessary to focus on the ideological characterization of users. 
Typically, obtaining data with a clear user's leaning toward a controversy is difficult. 
Hence the need to define a user's classification methodology to estimate their leaning about the debate according to the issue under analysis.

The task is modeled as a \textit{text classification} problem in the framework we are currently expanding. 
The raw text of posts and comments -- two features most OSNs have in common -- is encoded into a vector, which becomes the training set of a classification model specific to the context. 
This choice grants a higher level of generalizability while keeping the framework independent of any platform-specific feature, such as the number of \textit{likes} or \textit{retweets}, which are Twitter features often used by other frameworks. 
Among the various approaches to classification, the authors propose choosing between Deep Learning models or pre-trained Transformers while considering the amount of data and the type of information to be extracted. 
If it is necessary that the model keeps into account the semantic aspect of the sentence, the choice should fall on Transformers, otherwise, if it is just needed to extract specific information from the text may be also leveraged Neural Networks, e.g., LSTM and CNN. 
In addition, the classification may be modeled as a multiclass problem: thus, it is necessary to tackle the issue using a multiclass text classifier. 

\subsection*{Step 2: Modeling online debates}
One of the least considered dimensions by the echo chamber identification and analysis literature is time. 
Debates, both in online and offline arenas, unfold through time, and the stances of those individuals participating in them change as time goes by.
Therefore, it is important to longitudinally analyze the interactions that lead to the creation of polarized communities - e.g., echo chambers - in order to better understand system's dynamics effects.
Since ECs are epistemic systems in constant evolution, it is necessary to define an actionable way to model their topological counterpart to unveil interesting hidden patterns that may also help in their mitigation within OSNs. 

We model time-evolving interaction networks as a series of attributed snapshot graphs - extending the classical model defined in \cite{Greene2010}. 
Formally,

\begin{definition}[Attributed Snapshot Graph]
\begin{equation}
    \mathcal{G} = \langle G_1, G_2 ...G_t
 \rangle
\end{equation}
where each snapshot $G_i = (V_i, E_i, A_i)$ is a feature-rich graph univocally identified by the set of nodes $V_i$, edges $E_i$, and node labels $A_i$ - as described in Definition ~\ref{def:graph}.
\end{definition}
 
The timespan between adjacent snapshots is the key to good modeling of interactions: if too large, it is likely to have information loss in terms of varying nodes and links, if too short, the interaction graph may register only a few changes in the interactions, which may hide a possible temporal correlation with the process occurring in the network \cite{Caceres2011}. 
The criterion suggested by \cite{Salama2020} is to maintain a balance between the target to be studied and the temporal resolution.

\subsection*{Step 3: Identify Groups and their Dynamics}\label{sec:FrameworkECdetection}
So far, we defined our reference model for dynamic interactions involving agents enriched by some semantic information (e.g., their stance in a debate). 
The next step is to describe how to handle the extraction of meso-scale partitions inside the dynamic network through Dynamic Community Detection. 
As also suggested in the previous work, our choice for the partition extraction algorithm fell on EVA \cite{Citraro2020}, a Labeled Community Detection algorithm, applied on the graph representing each snapshot, which is able to optimize both structural cohesion and intra-community label homogeneity simultaneously. 
The algorithm extends the Louvain algorithm \cite{Blondel2008} to node-attributed graphs. 
On the one hand, it maximizes Newman's modularity and, on the other, a measure defined in its paper, known as \textit{Purity}. 
The two measures considered by EVA are defined as follows.

\begin{definition}[Modularity]
Modularity function quantifies the observed number of edges inside the given partition minus the expected number of edges if they were distributed following a null model of a random graph. 
The modularity has values ranging from -1 to +1. It is formalized as follows:
    \begin{equation}
    Q = \frac{1}{2m}\sum_{vw} \left[A_{uw} - \frac{k_v k_w}{(2m)}\right]\delta(c_v,c_w)
    \end{equation} 

\end{definition}

\begin{definition}[Purity]
Purity was defined in \cite{Citraro2020}, and it is calculated as the product of the frequencies of the most frequent labels carried by its node. This function lies within the range [0,1].
    \begin{equation}
       P_c=\prod_{a \in A} \frac{\max \left(\sum_{v \in c} a(v)\right)}{|c|}
    \end{equation}
\end{definition}

Optimizing both measures allows the extraction of partitions by simultaneously considering both modularity and the purity of the ideological clusters.

EVA is applied to every snapshot graph, and then, after extracting the partitions, ECs are detected by following the rationale in \cite{Morini2021}. 
The idea behind this is that polarized systems, especially ECs, need a sort of closed space in which an opinion can reverberate, moving from one member to another. 
Therefore, the idea suggested in \cite{Morini2021} consists of evaluating the partitions in terms of \textit{Conductance} and \textit{Purity}. The former has the function to estimate the number of edges volume staying inside the community, and the latter to assess the goodness of the partitions in terms of attribute homogeneity.

\begin{definition}[Conductance]
The conductance of a community $C$ is the volume of edges pointing out of it. 
The aim is to minimize the value of this function such that the average value across communities is as low as possible.
\begin{equation}
       Conductance_c = \frac{2\left|E_{O C}\right|}{2\left|E_C\right|+\left|E_{O C}\right|}
\end{equation}
where $E_{O C}$ is the number of edges exiting the community and $E_C$ is the number of edges remaining inside the community.
\end{definition}

According to these two measures, the risk of a community being an echo chamber is maximized when \textit{Conductance} is minimized -- it tends to $0$ -- and \textit{Purity} is maximized -- it tends to $1$. 
Following the original paper \cite{Morini2021}, we consider ECs, communities having, at the same time,  \textit{Conductance} equal to or less than $0.5$ and  \textit{Purity} equal to or greater than $0.7$. 
Notwithstanding, the two thresholds can be adjusted to fit the dataset under analysis better. 
Furthermore, we propose maintaining only the communities composed of at least $20$ users, thus removing small or noisy clusters.

Once identified the ECs, longitudinal analysis - aimed at assessing evolutive patterns - is performed by computing the pairwise Jaccard index among the communities of adjacent snapshots: an approach which has already been used in \cite{Greene2010} to identify the most likely evolution of partitions based on similarity. 
Before proceeding with this step, and to reduce the noise, we preprocessed the community sets, removing those users who joined online discussions by just posting/commenting once.
Moreover, we retained only the users it shares with adjacent ones for each snapshot, thus focusing on ``stable" sub-populations. 
We analyze the temporal development of ECs and non-polarized communities using a line plot in which each line represents the evolution of the similarity between adjacent partitions through timestamps. 
In addition, each line is enriched with a marker representing the \textit{status} of the community in a specific timestamp: triangles representing communities labeled as ECs, dots communities that are not. 
This type of plot allows, on the one hand, to assess the stability and evolution of \textit{individual} communities and, on the other, to observe the difference between all the ECs extracted using the approach previously described. 

\subsection*{Step 4: Topic extraction and analysis} 
After assessing the stability of ECs (and not-ECs) over time, we define a methodology to (i) capture the topics discussed by the identified clusters of users and (ii) compute the cluster-wide attitude towards that topic.
\\ \ \\
\textit{Topic modeling.} Among the various approaches, it was decided to employ an approach based on embeddings, i.e., BERTopic\footnote{BERTopic website: \url{https://maartengr.github.io/BERTopic/index.html}} \cite{Grootendorst2022}, which implements the BERT model for topic modeling tasks.
The motivations behind the choice are two: the nature of transformers, which allow for a better representation of words in context, and the competitive results of BERTopic \textit{w.r.t.} older topic modeling algorithms. 
In particular, BERTopic has the strength of being robust independently of the language model employed, despite performing or not performing the fine-tuning phase.
BERTopic extracts the topics in three steps. 
First, it generates the embedding of the input text using a language model and - to improve cluster quality - it reduces the data dimensions via UMAP \cite{McInnes2018} to avoid the curse of dimensionality \cite{Bellman1966}. 
Secondly, it clusters the embeddings through HDBSCAN \cite{McInnes2018}, which has the feature to consider noisy topics as outliers.
Finally, leveraging a class-based version of \textit{tf-idf} extracts the most meaningful words from each identified cluster.

To evaluate the quality of the obtained topics, we rely on two measures as proxies for an indicative -- and subjective -- human evaluation, as highlighted in \cite{Grootendorst2022}: namely, topic \emph{coherence}\cite{Grootendorst2022} and \emph{diversity} \cite{Dieng2020}. 
The former estimates the coherence of the extracted topics by using Normalized Pointwise Mutual Information (NPMI) \cite{Bouma2009}. 
It spans the range [-1, 1], with 1 a perfect association with scores given by human annotators.
The latter describes, for each topic, the percentage of unique words and lies within [0,1]. 
\\ \ \\
\textit{Valence analysis.}
Our aim is also to investigate the emotional component of ECs and discussions outside these closed systems. 
To address this issue, we rely on the VAD Lexicon and on KeyBERT \cite{Grootendorst2020keybert}, a method for keyphrase extraction that exploits BERT embeddings and cosine similarity to identify the most likely keywords describing a raw text. 
The main idea is to extract a set of keywords describing each post/comment and then proceed by calculating the valence score of the topic they are associated with. 
As a first step, keywords from the cleaned texts included in each topic are extracted. Secondly, the respective valence score in the lexicon is extracted for each keyword. 
The final valence score returned as output consists of the ratio between the sum of all the valence scores of the keywords found in the VAD lexicon and the total number of these keywords included in the lexicon. 
In this way, we wanted to reduce the noise in the results because of the presence of non-inherent words interfering with the score. 

\section*{Case study: Reddit socio-political dataset}\label{sec:CaseStudy}
In this section, we apply the proposed framework to a specific case study, discuss the obtained results, and evaluate its effectiveness and limitations.

The dataset we focus on is introduced in \cite{morini2020capturing,Morini2021} as the authors already assessed the presence of ECs, which we decided to track further from a temporal perspective. 
The dataset is composed of Reddit discussions about three socio-political topics. It focuses on the pro-/anti-Trump debate between January 2017 and June 2019, as it sharply exacerbated the clash between the two factions of Democrats and Republicans \cite{pewResearch2017}. 

Reddit is currently the seventh most used social network in the world \cite{redditStats}. It is particularly suitable as a source of data since it consists of \textit{subreddits}, topic-specific forums devoted to a single topic where users may freely discuss both general matters and more specific topics within various niches. 
Since the anonymity of the users is encouraged in these small forums, users may be motivated to talk more openly and, therefore, reach more extreme positions about controversial discussions, thus making Reddit a valuable source of data for the case study. 
\\
The data used for this case study are available on Github \footnote{Github repository with code and network/textual data: \url{https://github.com/ericacau/Trends-Topics_case_study}}.

\subsection*{Data collection and annotation}\label{sec:DataExtraction}
The three analyzed datasets are built on top of subreddits related to socio-political issues, categorized as follows: \textit{gun control}, \textit{minorities discrimination}, and \textit{politics}.
 \\
As a preliminary step, an additional dataset representing a \textit{polarized ground truth} was created - collecting posts and comments openly supporting or antagonizing Trump as president.
To such an extent, and to maintain a balanced representation of the two-sided controversy, post/comments data were extracted respectively from \textit{r/The\_Donald} and, \textit{r/Fuckthealtright}, \textit{r/EnoughTrumpSpam}. 
The resulting dataset was employed to train and test a classification model, namely $BERT_{BASE}$.
The model achieved an accuracy greater than 70\% for the test set \cite{Morini2021}.

The classifier was applied to the three Reddit sociopolitical datasets to infer, for each user, his/her leaning toward the specified controversy.
Each post/content was classified as either Pro- or Anti-Trump, and the prediction confidence (ranging in [0,1]) was used to assign a continuous value to the identified class.
Posts/comments with prediction confidence equal to 1 were considered perfectly aligned with a pro-Trump ideology, while the ones on the other extreme aligned with anti-Trump ones. 
Individual scores were subsequently averaged at the user level to compute the \textit{leaning score} 

$$L_u = \frac{\sum_{i=1}^n Prediction Score (p_i)}{n}$$
where $p_i \in P_i$ corresponds to a post shared by the user $u$ and $n = |P_i|$ indicates the cardinality of the set of posts published by the users. 
Finally, the resulting users' leaning scores values were discretized into intervals, as follows: if $L_u \leq 0.3$, then the posts are classified as \textit{antitrump}, as \textit{protrump} if $L_u \geq 0.7$ and \textit{neutral} otherwise. 
These thresholds, arbitrarily in nature, have also been maintained in this study but may be increased or decreased according to the dataset.
Further, Table ~\ref{tab:datasetStructure} shows the number of subreddits of the four datasets in terms of number of posts and users included.

\begin{table}[!ht]
\centering
\small
\caption{\textbf{Datasets statistics.}}
\begin{tabular}{|c|c|c|c|}
\hline
\textbf{Dataset}                   & \textbf{\# Subreddit} & \textbf{\# Post} & \textbf{\# User} \\ \hline
Ground Truth              & 3            & 302.762 & 68.113  \\
Gun control               & 6            & 180.170 & 65.111  \\
Minorities discrimination & 6            & 223.096 & 52.337  \\
Politics                  & 6            & 431.930 & 72.399 \\ \hline
\end{tabular}
\label{tab:datasetStructure}
\end{table}

\subsection*{Network modeling and EC identification}
\begin{figure}[!ht]
    \centering
    \caption{\textbf{Communities extracted in the five temporal snapshots. a) Gun control, b) Minorities discrimination, c) Politics. }}
    \subfloat[]{
        \includegraphics[scale=0.1]{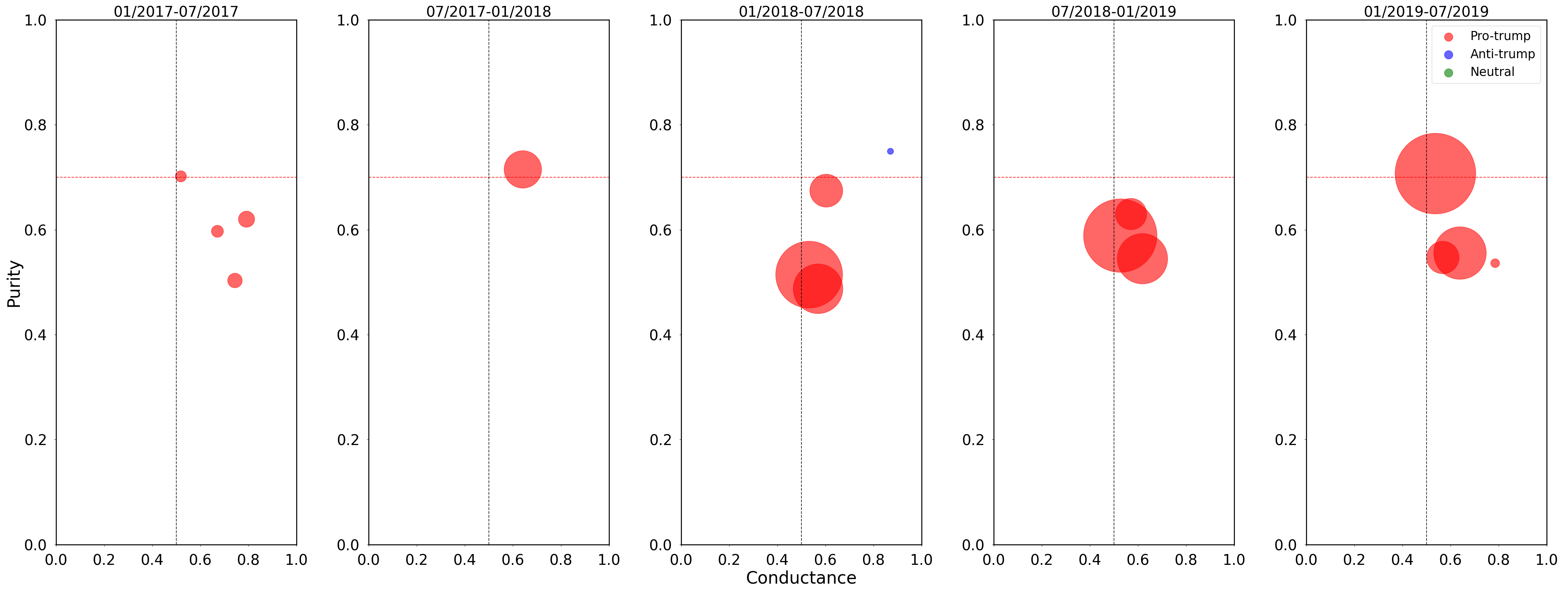}
        \label{fig:ECguncontrol}
    }
    \hfill
    \subfloat[]{
       \includegraphics[scale=0.1]{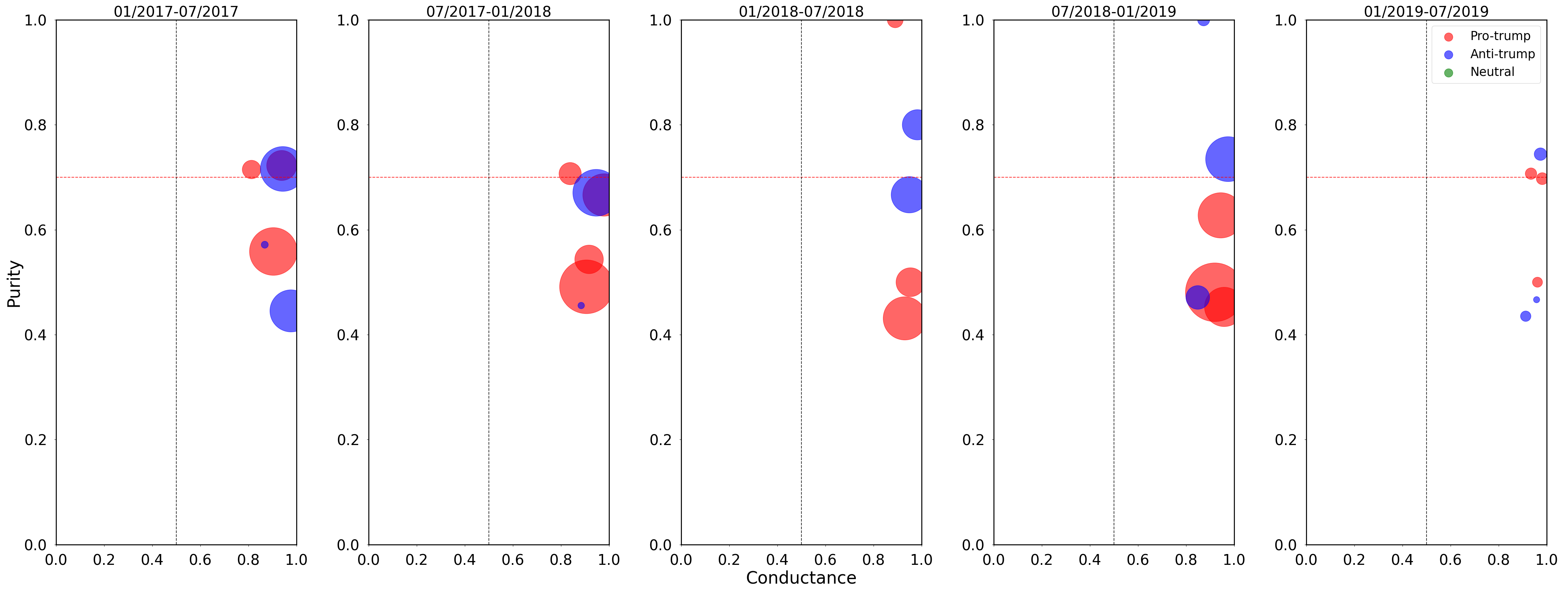}
        \label{fig:ECminority}
    }
    \hfill
    \subfloat[]{
        \includegraphics[scale=0.1]{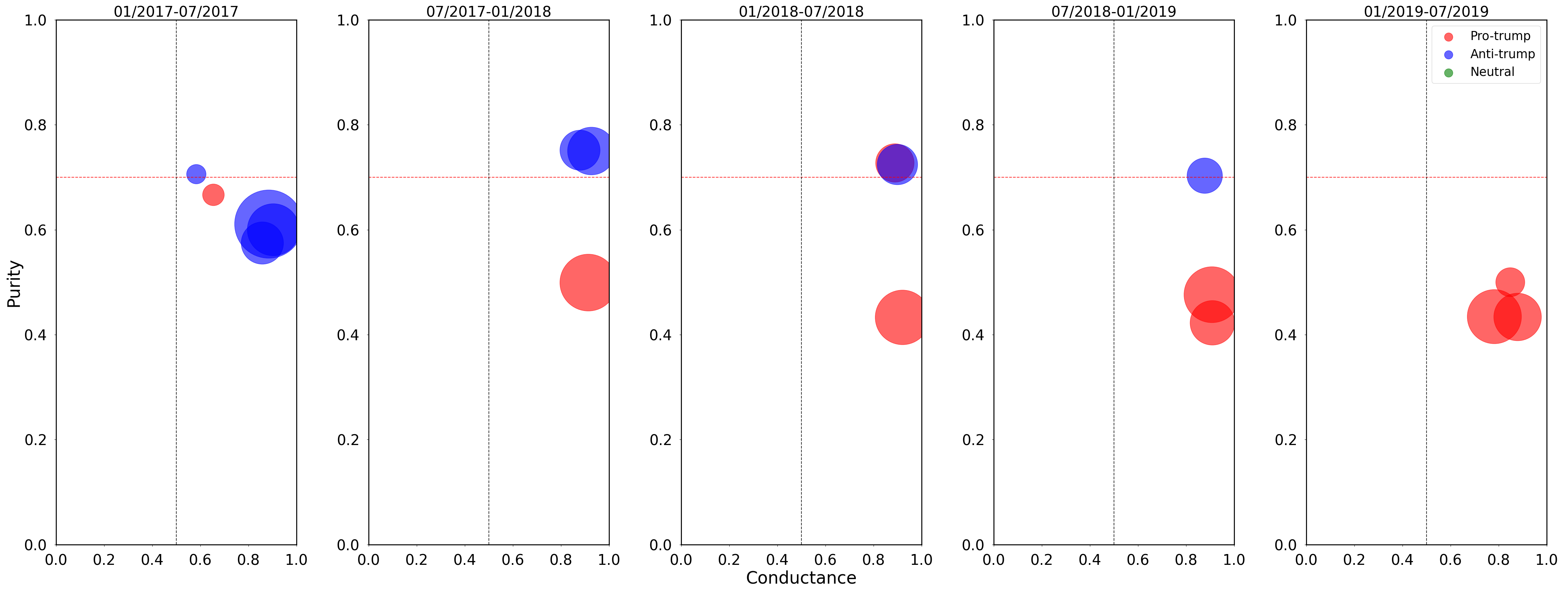}
        \label{fig:ECpolitics}
    }
    \label{fig:communitiesEC}
\end{figure}

Five snapshots were extracted from each of the three datasets, each covering a semester of the observed period. 
Starting from such a temporal discretization, a dynamic network was reconstructed as a snapshot sequence where a labeled user \textit{u} had an edge pointing towards user \textit{v} at time $t$, if and only if \textit{u} directly replied to a post or comment by user \textit{v} or vice versa during semester $t$. 
Each edge $(u, v, t)$ is then enriched with the weight of that tie, which is equal to the number of times the interaction between $u, v$ occurs during $t$.
Table ~\ref{tab:networkDescription} provides an overview of the network.
\begin{table}[]
\centering
\small
\caption{\textbf{Averaged number (per semester) of: nodes, edges, degree, and density of the networks, and distribution of neutral and pro-/anti-Trump nodes' leaning attribute.}}
\begin{tabular}{|c|c|c|c|}
\hline
\textbf{}                 & \textbf{Gun Control} & \textbf{\begin{tabular}[c]{@{}c@{}}Minorities\\ Discrimination\end{tabular}} & \textbf{Politics} \\ \hline
\textbf{\# Nodes}         & 11,388               & 6,617.6                            & 7,912.8           \\ \hline
\textbf{\# Edges}         & 114,262.4            & 80,497                             & 57,463.6          \\ \hline
\textbf{avg. degree}      & 17,1087              & 19,3608                            & 17,3634           \\ \hline
\textbf{avg. density}     & 0,00324              & 0,0031                             & 0,0013            \\ \hline
\textbf{Pro-Trump nodes}  & 2803.2               & 2150.4                             & 3837.6            \\ \hline
\textbf{Anti-Trump nodes} & 7385.2               & 3676,6                             & 2923.4            \\ \hline
\textbf{Neutral nodes}    & 1199.6               & 790,6                              & 1151.8            \\ \hline
\end{tabular}
\label{tab:networkDescription}
\end{table}
After network construction, communities were extracted from each snapshot through the Labeled Community Detection.
As previously discussed, the chosen algorithm was EVA because it can address both the optimization of modularity and label homogeneity.
The scatterplots in Figure ~\ref{fig:communitiesEC} show for \textit{Gun Control} (Figure ~\ref{fig:ECguncontrol}), \textit{Minorities discrimination} (Figure ~\ref{fig:ECminority}) and \textit{Politics} (Figure ~\ref{fig:ECpolitics}) the presence of polarized communities in each of the temporal snapshot.
The radius of the points in the scatterplot corresponds to the size of the community.
The vertical line describes the boundary set for Purity ($Purity>=0.7$), while the horizontal is the inverse of \textit{Conductance} ($1-Conductance >= 0.5$).

Ideally, the most polarized echo chambers are placed in the upper right corner in such graphical representations.
In addition, because of the temporal nature of this network, during each iteration from timestamp to timestamp, the Jaccard similarity index is applied to each identified partition. 
This led to the detection of the most likely evolution of a partition in $t$ at $t+1$.
The next step is detecting ECs through the Purity and Conductance thresholds. 
In this case study, we set the \textit{Conductance} score $\leq$ 0.5 and the \textit{Purity} $\geq$ 0.7 to ensure that most of the interactions in an EC involve users who stay within it while maintaining a high ideological cohesion.
\subsection*{Echo chambers' stability analysis}
\begin{figure}[!ht]
    \centering
    \caption{\textbf{Communities evolution through pairs of adjacent semesters. For each topic, the plot shows the value of the Jaccard index (\textit{y-axis}) through pair of adjacent semesters (\textit{x-axis}). Triangles mark the community as an echo chamber, while circles as not echo chambers. a) Gun Control, b) Minorities discrimination, c) Politics.}}
    \subfloat[]{
        \includegraphics[scale=0.3]{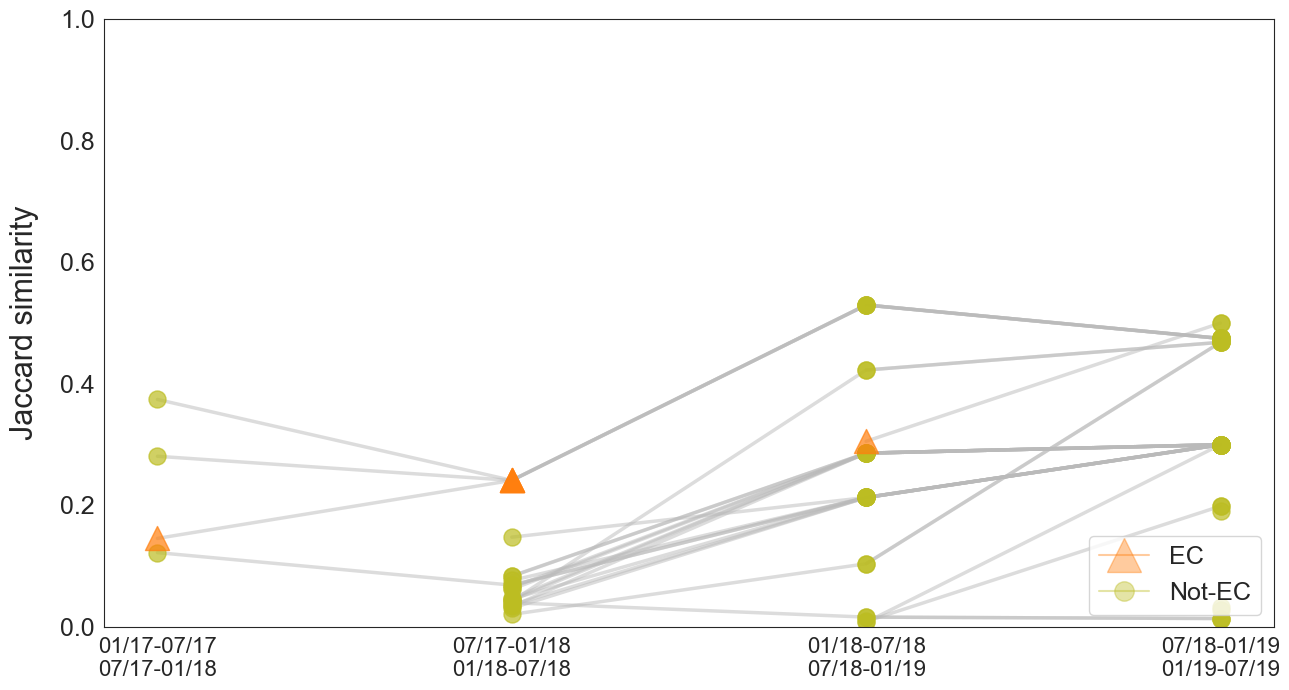}
    \label{fig:stability_guncontrol}
    }
    \hfill
    \subfloat[]{
       \includegraphics[scale=0.3]{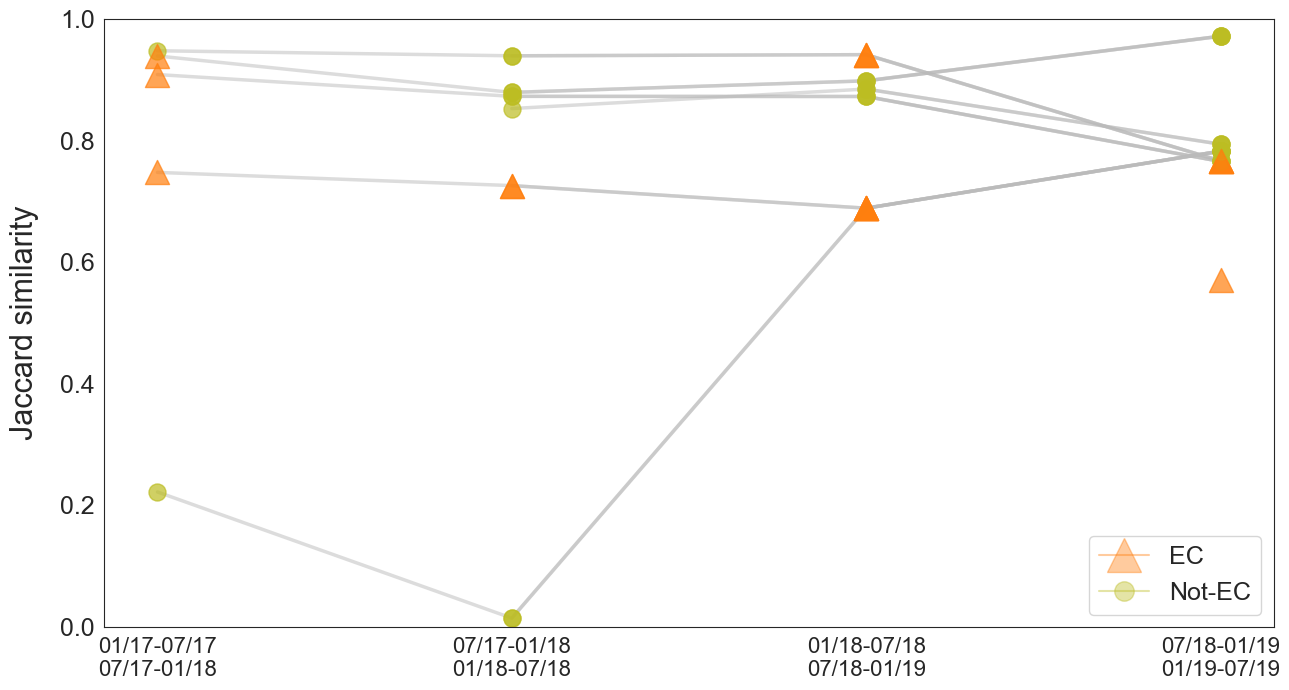}
        \label{fig:stability_minority}
    }
    \hfill
    \subfloat[]{
        \includegraphics[scale=0.3]{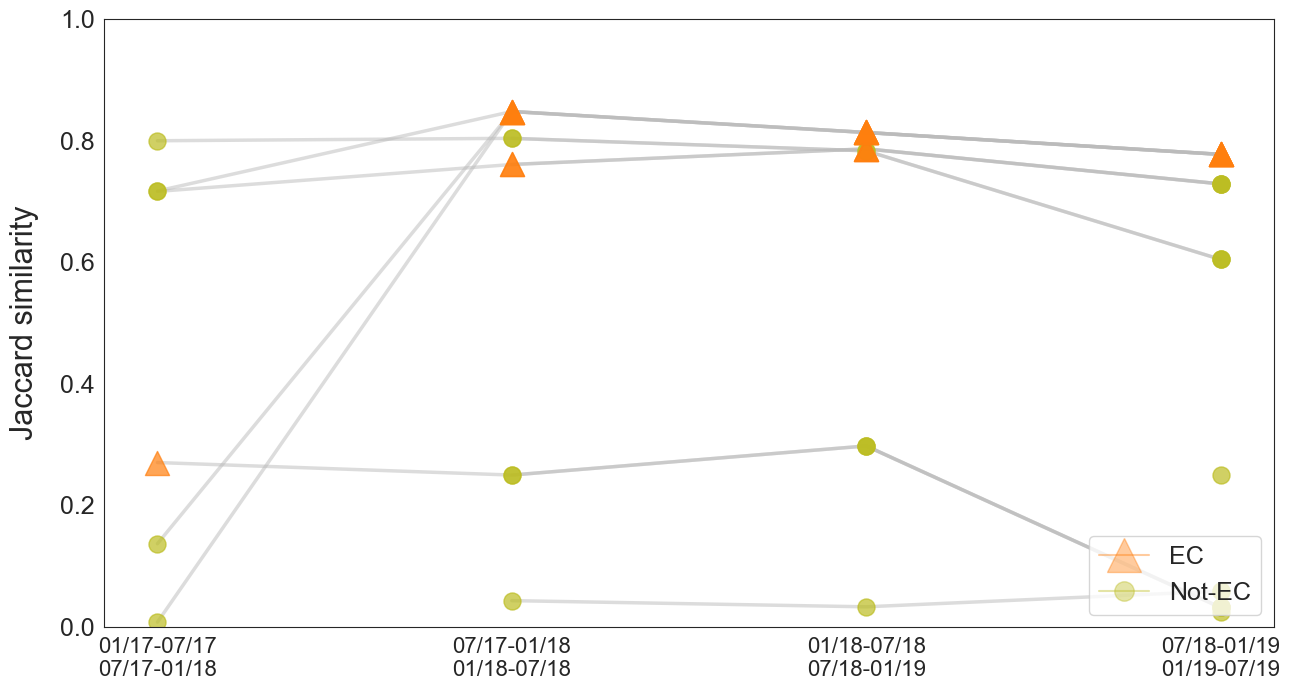}
        \label{fig:stability_politics}
    }
    \label{fig:fig:stability_plot}
\end{figure}

In the previous section, we assessed the presence of ECs in every static snapshot of the network. 
The analysis now moves toward understanding ECs' internal dynamics to answer two research questions. 
\begin{itemize}
    \item RQ1: \textit{Are echo chambers stable over time w.r.t. the users that compose them?}
    \item RQ2: \textit{Do echo chambers keep or lose their polarization as time passes?}
\end{itemize}
The results, shown in Fig. ~\ref{fig:stability_plot}, differed slightly for each of the three main categories of discussions analyzed. 
It should be noted that the ECs belonging to \textit{Politics} and \textit{Minorities discrimination} appear to be stable even over a long timespan, with variations that may be ascribed to the different topics discussed and to the temporal segmentation chosen for the snapshots.
Instead, ECs in \textit{Gun control} behave differently than those in the other two topics.
Fig. ~\ref{fig:stability_guncontrol} shows the evolution of detected ECs. 
In this case, the overall Jaccard similarity between adjacent timestamps is very low. 
The only exception is an EC that turns into a lesser polarized community between the end of 2017 and the beginning of 2018, which reaches a similarity value equal to 0.53, against the 0.24 of the previous pair of semesters.

This behavior is less pronounced in \textit{Minorities discrimination} and \textit{Politics} (Fig. ~\ref{fig:stability_minority}, ~\ref{fig:stability_politics}). 
First, in both cases, the internal stability is very high from the beginning of monitoring, reaching its highest value, 93\%, during the first year of discussions about \textit{Minorities discrimination}. 
Then, as time passes, the internal similarity seems to decrease slightly, and in certain cases, ECs become communities that do not bear a strong ideological cohesion as in the previous pair of adjacent semesters. 
In other cases, otherwise, they remain the same. 
In \textit{Minorities discrimination} (Fig. ~\ref{fig:stability_minority}), for example, it is interesting to note that the EC with the lowest percentage of common users between the first two semesters, namely 75\%, is also the one with the longest lifecycle as EC, as it maintains its status until the very last pair of analyzed snapshots. 
Such behavior might also be identified in \textit{Politics} (Fig. ~\ref{fig:stability_politics}), which has a similar case of an EC that took root in the second semester of 2017 but stayed almost still until the end of the monitoring. 
Furthermore, seems that the less polarized communities derived from ECs do not become ECs again.
In addition, in the discussed figures, there is the presence of ECs persisting over a shorter but still significant timespan of only six months.

Regarding Not-ECs, from Fig. ~\ref{fig:stability_guncontrol}, it can be noted that \textit{Gun control} communities experience an increase in their internal stability between the second and the third semester, without reaching strong stability as the one characterizing ECs. Diversely in both \textit{Minorities discrimination} and \textit{Politics}, communities' internal stability tends to be very high except for one community in the first dataset and two communities in the second. Furthermore, they may become even more stable if the debate becomes more heated, which leads them to evolve into echo chambers.

\subsection*{Topic modeling on polarized systems}\label{sec:topicModeling}
After observing the overall persistence over time and the intrinsic stability of most ECs in terms of user compositions, we inspect the textual productions made by the users. 

The first step before applying the algorithm was to create, for each topic and semester, a dataset containing the texts produced by users and information about users themselves, including their being or not being members of ECs. 
This was necessary since our goal is to characterize and distinguish between text produced inside ECs and non-polarized communities.
Therefore, we applied BERTopic on all the documents (i.e., covering texts produced inside ECs and not-ECs users), thus identifying 13 topics. 
The corpus of textual data on which the model was trained was preprocessed as follows to clean the raw text in input and the resulting labels in output. 
First, the preprocessing involved the normalization and cleaning of the raw text. 
The cleaning step included expanding abbreviations and short forms and removing markdown characters. 
In addition, as a preprocessing step, words were lemmatized using Wordnet \cite{Fellbaum1998}. 
Finally, a set of representative keywords of the entire text was extracted for each post of the dataset using KeyBERT \cite{Grootendorst2020keybert}, thus creating a fine-tuned vocabulary to label the identified topics better.
To reduce the number of outliers originally identified by BERTopic on the available data, we decided to substitute the default clustering algorithm it employs (HDBSCAN \cite{McInnes2018}) with K-Means \cite{Lloyd1982}. 
The minimum clusters size was set to 120 to avoid smaller -- and noisier -- topics. 
Moreover, the clusters were further diversified by employing the \textit{Maximum Marginal Relevance} \cite{Carbonell1998} ranking algorithm, which allowed the identification of the most meaningful and diverse words describing each topic.

\begin{table}[!h]
\centering
\caption{\textbf{Comparison between LDA (13 topics) and BERTopic in terms of topic coherence and diversity.} }
\begin{tabular}{|c|cc|cc|}
\hline
                                                                             & \multicolumn{2}{c|}{\textbf{Topic coherence}}         & \multicolumn{2}{c|}{\textbf{Topic diversity}}         \\ \hline
                                                                             & \multicolumn{1}{c|}{\textbf{LDA}} & \textbf{BERTopic} & \multicolumn{1}{c|}{\textbf{LDA}} & \textbf{BERTopic} \\ \hline
\textbf{Guncontrol}   
& \multicolumn{1}{c|}{0.0222}       & 0.1661            & \multicolumn{1}{c|}{0.923}             &           0.8376        \\ 
\textbf{\begin{tabular}[c]{@{}c@{}}Minorities\\ discrimination\end{tabular}} & \multicolumn{1}{c|}{0.0082}       & 0.2533            & \multicolumn{1}{c|}{0.9487}         &   0.9145                \\ 
\textbf{Politics}                                                            & \multicolumn{1}{c|}{-0.009}       & 0.1997            & \multicolumn{1}{c|}{0.9487}             &       0.9316            \\ \hline
\end{tabular}
\label{tab:topic_coherence_diversity}
\end{table}
As for the results of topic modeling (see Table ~\ref{tab:topic_coherence_diversity} for \textit{Coherence} and \textit{Diversity} values), in both ECs and less polarized communities belonging to the \textit{\textit{Gun control}} dataset, it was possible to assess the overall presence of generic discussions about guns and ammunition brands as well as requests for advice from expert users.
The broadness of the topics leaves small room for other topics related to real-world events.
Interestingly, at the beginning of 2018, one EC focused mainly on a particularly controversial topic: the War in Syria. In the following semester, users discussed instead the 2018 Firearms Amendment Act; then, the focus shifted back to the War in Syria.

Users falling inside the \textit{Minorities discrimination} dataset discussed a wider variety of topics.
One interesting issue that emerged is Gamergate, an online social movement, for which one of the subreddits included in the dataset, namely \textit{r/KotakuInAction}, represents \textit{the main hub} on Reddit, as stated on the subreddit homepage.  
The campaign started in 2014 to harass female journalists and developers involved in the video game industry who experienced doxing, rape, and death threats \cite{Jhaver2018}.
Despite the lack of leaders or internal organization, it rapidly evolved into a broader movement targeting \textit{Social Justice Warrior} \footnote{"Social Justice Warrior", Cambridge Dictionary. Accessed July 18, 2023. Available: \url{https://dictionary.cambridge.org/dictionary/english/social-justice-warrior}} activists and the perceived excess of political correctness in video games. 
According to Massanari \cite{Massanari2016}, who addressed the movement as an "echo chamber of anger", comprises people sharing the same core values of toxic masculine gaming culture, who may see the presence of women in the game industry as a threat.
In addition, it has also been addressed as ideologically near to the alt-right wing of the political spectrum \cite{Massanari2016}.

Other controversial issues that emerged are \textit{antifascists} movements, often discussed along with the protests in Berkeley during 2017.
Events started in 2017 when the right-wing supporter Milo Yiannapoulos was invited as a speaker at UC Berkeley.
He encountered opposition from a group of armed anti-fascists who turned regular student opposition into a violent riot, damaging the university infrastructure and assaulting police forces. 
This first event turned into a chain of violent events that culminated at the end of August in the \textit{Rally Against Hate}, in which far-left protesters clashed with far-right supporters. 

Outside ECs, other recurring discussions were focused on more general but still polarized topics, e.g., the \textit{white privilege}, Canadian politics, and gender equality.

Inside \textit{Politics}, users belonging to ECs mainly discussed abortion and the \textit{Mueller special counsel investigation}, conducted to assess the interference of Russia in the 2016 U.S. elections.
For example, the second topic was the main focus of the discussion in the echo chamber with the longest lifecycle shown in Fig. ~\ref{fig:stability_politics}.
Outside ECs, the communities seem to discuss other issues besides those discussed in ECs, e.g., news about Trump and politics, Obamacare, and Libertarianism.   

Further details on topic modeling can be found in the Supplementary Materials.
\subsection*{Valence analysis}

After assessing the presence of meaningful and polarized topics across all three macro-topics, the analysis deepened to gain insights into the pleasantness or unpleasantness of the topics that emerged from the words chosen by users. 
This emotive attitude of users was investigated using the NRC-Valence Lexicon \cite{Mohammad2018}, which allowed for extracting the average Valence score from all the posts included in the macro-category represented by a topic. 
Specifically, for each topic, the average valence was calculated as the ratio of the sum of the valences of its keywords that are annotated in the VAD \textit{w.r.t.} the total number of keywords in the lexicon. 
The results were then visually investigated with the aid of a scatterplot where each point is colored according to the average valence score, i.e., orange for negativeness, pearl white for neutral, and blue for positiveness. 

The results reflect the inherently polarized nature of the topics under analysis, with only slight differences between ECs and communities.

\begin{figure}[!ht]
    \centering
    \caption{\textbf{Topic valence (\textit{x-axis}) for EC and Not-ECs (\textit{y-axis}) users' clusters. Colors describe the attitudes conveyed in texts. Strongly polarized topics are characterized by a blue or dark orange hue, the former for positive and the latter for negatively connotated topics. Circle sizes relate to the number of texts associated with each topic. a)\textit{ Gun control}, b) \textit{Minorities discrimination}, c) \textit{Politics}}}
    \subfloat[]{
        \includegraphics[scale=0.5]{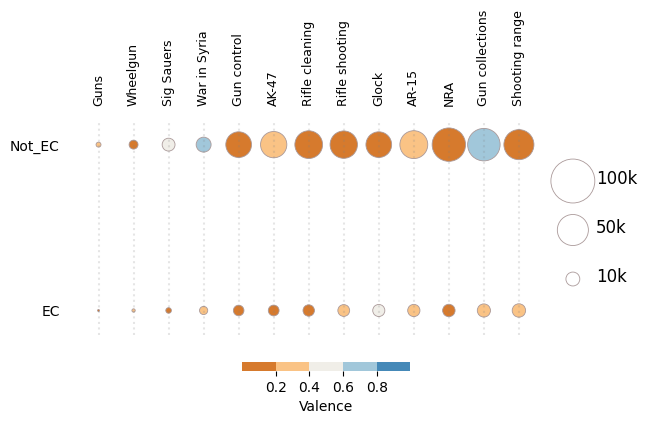}
        \label{fig:valence_guncontrol}
    }
    \hfill
    \subfloat[]{
       \includegraphics[scale=0.5]{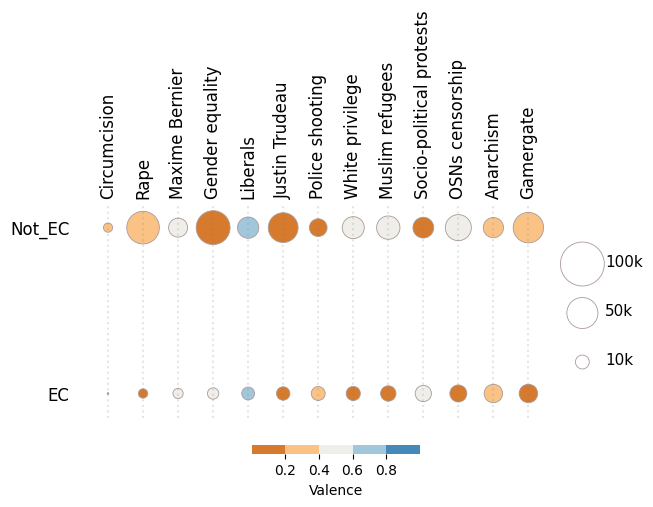}
        \label{fig:valence_minority}
    }
    \hfill
    \subfloat[]{
        \includegraphics[scale=0.5]{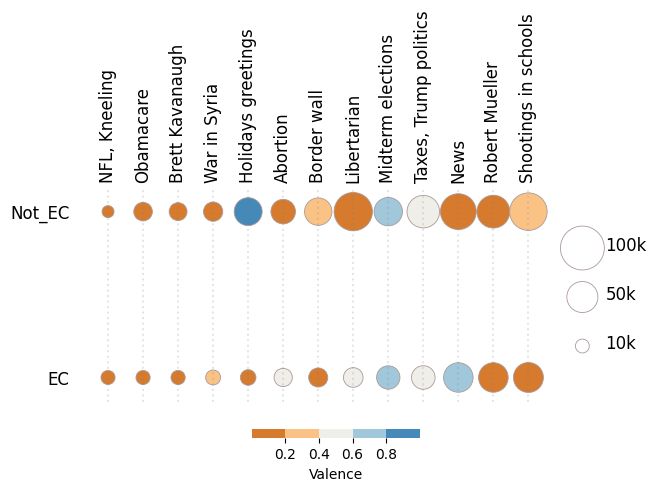}
        \label{fig:valence_politics}
    }
    \label{fig:stability_plot}
\end{figure}

In \textit{Gun control}, Fig. ~\ref{fig:valence_guncontrol}, the difference between the two systems is not as stark as we expected, as users outside ECs appear to discuss using more negative words than users inside ECs, except when talking specifically about \textit{Gun collections}, where users outside ECs often use terms associated with negative meaning. 

Conversely, in \textit{Minorities discrimination}, Fig. ~\ref{fig:valence_minority}, it is worth noting that ECs users discuss only one topic with a positive attitude, i.e., discussions about the center-left wing of the political spectrum. 
Moreover, we can also observe how topics that tend to be strongly negatively polarized outside ECs - e.g., fascism, racism, and police shootings against minorities - are more neutral inside ECs. 
Such a result, which might appear contradictory at first glance, may be related to the fact that in ECs, users tend to have less negative or condemning opinions on fascism and sensitive issues. 
Furthermore, the Gamergate controversy is characterized by an increase in wording negativeness, which the misogynistic nature of the movement may justify.
Thus, we can assume that users are prone to condemn and attack women and minorities using negatively connoted language.
This result is reflected and magnified by an ever more polarized negative attitude about the topic "OSNs censorship", representing another of the core arguments discussed by GamerGate supporters against SJWs. 
In particular, the topic refers to the Twitter ban on journalists involved in the movement, including Milo Yiannapoulos.


Focusing on the topics identified in the \textit{Politics} dataset, in Fig. ~\ref{fig:valence_politics}, emerges a more negative connotation in discussions about school shootings and protests against them. 
Moreover, similarly to what was observed in the \emph{Minority} dataset, a subset of topics are treated as less negative in ECs w.r.t. not-ECs - e.g.,  War in Syria and abortion. 
At the same time, a more negative attitude emerges toward the border wall between Mexico and the United States.

Despite these results, however, it can be argued that the average valence score alone seems insufficient to highlight a clear distinction in terms of sentiment between ECs and Not-ECs. 

\section*{Conclusions}
In this work, we proposed a platform-independent framework to capture the dynamics of ECs and inspect the content and the positiveness or negativeness conveyed in their discussions. 
The framework is composed of four steps and leverages only posts and comments, a feature common to most OSNs that represents both the topology of relations and the analyzed content. 
As a result, the framework turns out to be highly reusable in other OSNs different from Reddit, as long as a system of posts and comments is implemented in the platform. 

The framework has the advantage of considering one of the pillars in common with most ECs definitions, namely, the idea of being closed systems of like-minded users mainly interacting with one another. 
This is obtained via ECs extraction from a meso-scale topological level and node-attributed snapshot graph, allowing for a simpler, but at the same time, effective representation of the dynamics of relations over time. 
Furthermore, it considers another crucial component that makes ECs a \textit{polarized system}, the ideology or leaning assumed by users during discussions.

From the case study we presented, we observed different tendencies, but the main one seems to be that ECs keep trapped a large portion of users, while a small component leaves the polarized space as time passes. 
The most interesting result was observed in the \textit{Politics} and \textit{Minorities discrimination} discussions, as they were characterized by ECs with high stability for almost two years. 

Topic modeling allowed for the extraction of a wide range of interpretable discussion themes. 
Topics analysis was then enriched with emotion analysis by extracting scores describing the valence of the texts included in each topic, thus leading to particularly interesting observations. 
First, from the visual analysis of the results, it was possible to infer that in ECs, most topics were discussed using words conveying a negative meaning. 
Secondly, interestingly, we were able to observe that often, when highly divisive topics are taken into account - i.e., racism - users trapped within ECs tend to use neutral wording, while outside ECs, a more negative one.

\subsection*{Approach weaknesses and limitations}
The proposed framework has weaknesses and limitations that must be discussed and considered.
Firstly, the echo chamber concept is something on which there is currently no consensus, even from a qualitative perspective. 
Similarly, the core concept of the topological extraction of ECs in this framework, namely the notion of \textit{community detection}, is well-known in the network science literature because it is an \textit{ill-posed problem}.
Secondly, the framework generally lacks rigorous validation related to the absence of ground truth for labels describing the user leaning. 
These labels are inferred through a classifier trained on polarized ground truth and act as a mere \textit{proxy} to understand people's real -- and multi-faceted -- political leaning. 
Even topic modeling results suffer from a similar problem as they employed an unsupervised approach for which the extracted topics were not annotated.

Finally, an intrinsic limitation is due to the stochastic nature of the UMAP employed by BERTopic. 

\subsection*{Future developments}
To better understand the complex nature of ECs, we plan to perform a more in-depth analysis of both network topology and textual data produced by users.

In particular, we plan to move from pairwise to high-order interactions, thus explicitly accounting for group dynamics. 
In this way, we might be able to capture a wider range of interactions that might provide insight into homophilic behaviors related to the phenomenon, e.g., peer pressure.

Furthermore, we aim to enhance content analysis by integrating and studying the \textit{stance} of users towards the controversy in which ECs are detected. 
Stance detection is an NLP problem that fits well with the concept of echo chambers because it is related to the prediction of users' ideas (pro, none, or against) toward a target \cite{Alturayeif2023}. 

Finally, the proposed framework will be applied to other case studies - both on controversial and less contentious issues - to properly observe whether similar patterns characterizing ECs are found even outside polarized discussions. 
For example, testing the framework on other case studies comparing polarizing issues and neutral topics would allow us to verify the hypothesis about the higher neutrality used when discussing controversial issues inside ECs. 
This would also increase the number of studies on the dynamic development of ECs, given their scarcity in literature.

\section*{Acknowledgment}
This work is supported by: the EU NextGenerationEU programme under the funding schemes PNRR-PE-AI FAIR (Future Artificial Intelligence Research); the EU – Horizon 2020 Program under the scheme ``INFRAIA-01-2018-2019 – Integrating Activities for Advanced Communities” (G.A. n.871042) ``SoBigData++: European Integrated Infrastructure for Social Mining and Big Data Analytics” (http://www.sobigdata.eu); PNRR-``SoBigData.it - Strengthening the Italian RI for Social Mining and Big Data Analytics'' - Prot. IR0000013;

\bibliography{bibliography.bib}

\begin{thebibliography}{10}

\bibitem{Floridi2014}
Luciano Floridi.
\newblock {\em The Fourth Revolution: How the Infosphere is Reshaping Human Reality}.
\newblock Oxford University Press UK, 2014.

\bibitem{Festinger1962}
Leon Festinger.
\newblock Cognitive dissonance.
\newblock {\em Scientific American}, 207(4):93--106, October 1962.

\bibitem{Morini2021}
Virginia Morini, Laura Pollacci, and Giulio Rossetti.
\newblock Toward a standard approach for echo chamber detection: Reddit case study.
\newblock {\em Applied Sciences}, 11(12):5390, June 2021.

\bibitem{Conover2021}
Michael Conover, Jacob Ratkiewicz, Matthew Francisco, Bruno Goncalves, Filippo Menczer, and Alessandro Flammini.
\newblock Political polarization on twitter.
\newblock {\em Proceedings of the International {AAAI} Conference on Web and Social Media}, 5(1):89--96, August 2021.

\bibitem{Adamic2005}
Lada~A. Adamic and Natalie Glance.
\newblock The political blogosphere and the 2004 u.s. election: divided they blog.
\newblock In {\em Proceedings of the 3rd international workshop on Link discovery}, KDD05, page 36–43. ACM, August 2005.

\bibitem{Guerra2021}
Pedro Guerra, Wagner~Meira Jr., Claire Cardie, and Robert Kleinberg.
\newblock A measure of polarization on social media networks based on community boundaries.
\newblock {\em Proceedings of the International {AAAI} Conference on Web and Social Media}, 7(1):215--224, August 2021.

\bibitem{Edwards2013}
Arthur Edwards.
\newblock (how) do participants in online discussion forums create `echo chambers'?
\newblock {\em Argumentation in political deliberation}, 2(1):127--150, May 2013.

\bibitem{Gilbert2009}
E.~Gilbert, T.~Bergstrom, and K.~Karahalios.
\newblock Blogs are echo chambers: Blogs are echo chambers.
\newblock In {\em 2009 42nd Hawaii International Conference on System Sciences}, pages 1--10, 2009.

\bibitem{Ge2020}
Yingqiang Ge, Shuya Zhao, Honglu Zhou, Changhua Pei, Fei Sun, Wenwu Ou, and Yongfeng Zhang.
\newblock Understanding echo chambers in e-commerce recommender systems.
\newblock In {\em Proceedings of the 43rd International ACM SIGIR Conference on Research and Development in Information Retrieval}, SIGIR ’20, page 2261–2270. ACM, July 2020.

\bibitem{An2014}
Jisun An, Daniele Quercia, and Jon Crowcroft.
\newblock Partisan sharing.
\newblock In {\em Proceedings of the second {ACM} conference on Online social networks}, pages 13--24. {ACM}, October 2014.

\bibitem{Bakshy2015}
Eytan Bakshy, Solomon Messing, and Lada~A. Adamic.
\newblock Exposure to ideologically diverse news and opinion on facebook.
\newblock {\em Science}, 348(6239):1130--1132, June 2015.

\bibitem{Caldern2019}
Fernando~H. Calder{\'{o}}n, Li-Kai Cheng, Ming-Jen Lin, Yen-Hao Huang, and Yi-Shin Chen.
\newblock Content-based echo chamber detection on social media platforms.
\newblock In {\em Proceedings of the 2019 {IEEE}/{ACM} International Conference on Advances in Social Networks Analysis and Mining}, pages 597--600. {ACM}, August 2019.

\bibitem{Garimella2018}
Kiran Garimella, Gianmarco De~Francisci Morales, Aristides Gionis, and Michael Mathioudakis.
\newblock Political discourse on social media.
\newblock In {\em Proceedings of the 2018 World Wide Web Conference on World Wide Web - {WWW} {\textquotesingle}18}, page 913–922. {ACM} Press, 2018.

\bibitem{Kratzke2023}
Nane Kratzke.
\newblock How to find orchestrated trolls? a case study on identifying polarized twitter echo chambers.
\newblock {\em Computers}, 12(3):57, March 2023.

\bibitem{DeFrancisciMorales2021}
Gianmarco De~Francisci Morales, Corrado Monti, and Michele Starnini.
\newblock No echo in the chambers of political interactions on reddit.
\newblock {\em Scientific Reports}, 11(1), February 2021.

\bibitem{Villa2021}
Giacomo Villa, Gabriella Pasi, and Marco Viviani.
\newblock Echo chamber detection and analysis.
\newblock {\em Social Network Analysis and Mining}, 11(1), August 2021.

\bibitem{Cinelli2021}
Matteo Cinelli, Gianmarco De~Francisci Morales, Alessandro Galeazzi, Walter Quattrociocchi, and Michele Starnini.
\newblock The echo chamber effect on social media.
\newblock {\em Proceedings of the National Academy of Sciences}, 118(9), February 2021.

\bibitem{Palla2007}
Gergely Palla, Albert-L{\'{a}}szl{\'{o}} Barab{\'{a}}si, and Tam{\'{a}}s Vicsek.
\newblock Quantifying social group evolution.
\newblock {\em Nature}, 446(7136):664--667, April 2007.

\bibitem{Cazabet2019}
Remy Cazabet and Giulio Rossetti.
\newblock Challenges in community discovery on temporal networks.
\newblock In {\em Computational Social Sciences}, pages 181--197. Springer International Publishing, 2019.

\bibitem{Kopacheva2022}
Elizaveta Kopacheva and Victoria Yantseva.
\newblock Users' polarisation in dynamic discussion networks: The case of refugee crisis in sweden.
\newblock {\em {PLOS} {ONE}}, 17(2):e0262992, February 2022.

\bibitem{Liu2016}
Lin Liu, Lin Tang, Wen Dong, Shaowen Yao, and Wei Zhou.
\newblock An overview of topic modeling and its current applications in bioinformatics.
\newblock {\em {SpringerPlus}}, 5(1), September 2016.

\bibitem{Hospedales2011}
Timothy Hospedales, Shaogang Gong, and Tao Xiang.
\newblock Video behaviour mining using a dynamic topic model.
\newblock {\em International Journal of Computer Vision}, 98(3):303--323, December 2011.

\bibitem{Blei2003}
David~M. Blei, Andrew~Y. Ng, and Michael~I. Jordan.
\newblock Latent dirichlet allocation.
\newblock {\em J. Mach. Learn. Res.}, 3(null):993–1022, mar 2003.

\bibitem{Moody2016}
Christopher~E Moody.
\newblock Mixing dirichlet topic models and word embeddings to make lda2vec, 2016.

\bibitem{Mikolov2013}
Tomas Mikolov, Ilya Sutskever, Kai Chen, Greg Corrado, and Jeffrey Dean.
\newblock Distributed representations of words and phrases and their compositionality.
\newblock In {\em Proceedings of the 26th International Conference on Neural Information Processing Systems - Volume 2}, NIPS'13, page 3111–3119, Red Hook, NY, USA, 2013. Curran Associates Inc.

\bibitem{Blei2006}
David~M. Blei and John~D. Lafferty.
\newblock Dynamic topic models.
\newblock In {\em Proceedings of the 23rd international conference on Machine learning - {ICML} {\textquotesingle}06}, page 113–120. {ACM} Press, 2006.

\bibitem{Iwata2009}
Tomoharu Iwata, Shinji Watanabe, Takeshi Yamada, and Naonori Ueda.
\newblock Topic tracking model for analyzing consumer purchase behavior.
\newblock In {\em Proceedings of the 21st International Joint Conference on Artificial Intelligence}, IJCAI'09, page 1427–1432, San Francisco, CA, USA, 2009. Morgan Kaufmann Publishers Inc.

\bibitem{Devlin2018}
Jacob Devlin, Ming-Wei Chang, Kenton Lee, and Kristina Toutanova.
\newblock Bert: Pre-training of deep bidirectional transformers for language understanding, 2018.

\bibitem{Liu2019}
Yinhan Liu, Myle Ott, Naman Goyal, Jingfei Du, Mandar Joshi, Danqi Chen, Omer Levy, Mike Lewis, Luke Zettlemoyer, and Veselin Stoyanov.
\newblock Roberta: A robustly optimized bert pretraining approach.
\newblock {\em ArXiv}, abs/1907.11692, 2019.

\bibitem{Lewis2020}
Mike Lewis, Yinhan Liu, Naman Goyal, Marjan Ghazvininejad, Abdelrahman Mohamed, Omer Levy, Veselin Stoyanov, and Luke Zettlemoyer.
\newblock {BART}: Denoising sequence-to-sequence pre-training for natural language generation, translation, and comprehension.
\newblock In {\em Proceedings of the 58th Annual Meeting of the Association for Computational Linguistics}, pages 7871--7880, Online, July 2020. Association for Computational Linguistics.

\bibitem{Sanh2019}
Victor Sanh, Lysandre Debut, Julien Chaumond, and Thomas Wolf.
\newblock Distilbert, a distilled version of bert: smaller, faster, cheaper and lighter.
\newblock {\em ArXiv}, abs/1910.01108, 2019.

\bibitem{Vaswani2017}
Ashish Vaswani, Noam Shazeer, Niki Parmar, Jakob Uszkoreit, Llion Jones, Aidan~N. Gomez, \L{}ukasz Kaiser, and Illia Polosukhin.
\newblock Attention is all you need.
\newblock In {\em Proceedings of the 31st International Conference on Neural Information Processing Systems}, NIPS'17, page 6000–6010, Red Hook, NY, USA, 2017. Curran Associates Inc.

\bibitem{Russell2003}
James~A. Russell.
\newblock Core affect and the psychological construction of emotion.
\newblock {\em Psychological Review}, 110(1):145--172, 2003.

\bibitem{Bradley1999}
Margaret~M. Bradley and Peter~J. Lang.
\newblock Affective norms for english words (anew): Instruction manual and affective ratings.
\newblock Technical report, 1999.

\bibitem{Warriner2013}
Amy~Beth Warriner, Victor Kuperman, and Marc Brysbaert.
\newblock Norms of valence, arousal, and dominance for 13, 915 english lemmas.
\newblock {\em Behavior Research Methods}, 45(4):1191--1207, February 2013.

\bibitem{Mohammad2018}
Saif Mohammad.
\newblock Obtaining reliable human ratings of valence, arousal, and dominance for 20,000 {E}nglish words.
\newblock In {\em Proceedings of the 56th Annual Meeting of the Association for Computational Linguistics (Volume 1: Long Papers)}, pages 174--184, Melbourne, Australia, July 2018. Association for Computational Linguistics.

\bibitem{Louviere2015}
Jordan~J. Louviere, Terry~N. Flynn, and A.~A.~J. Marley.
\newblock {\em Best-Worst Scaling}.
\newblock Cambridge University Press, September 2015.

\bibitem{Greene2010}
Derek Greene, D{\'{o}}nal Doyle, and P{\'{a}}draig Cunningham.
\newblock Tracking the evolution of communities in dynamic social networks.
\newblock In {\em 2010 International Conference on Advances in Social Networks Analysis and Mining}, pages 176--183. {IEEE}, August 2010.

\bibitem{Caceres2011}
Rajmonda~Sulo Caceres, Tanya Berger-Wolf, and Robert Grossman.
\newblock Temporal scale of processes in dynamic networks.
\newblock In {\em 2011 {IEEE} 11th International Conference on Data Mining Workshops}, pages 925--932. {IEEE}, December 2011.

\bibitem{Salama2020}
Mohamed Salama, Mohamed Ezzeldin, Wael El-Dakhakhni, and Michael Tait.
\newblock Temporal networks: a review and opportunities for infrastructure simulation.
\newblock {\em Sustainable and Resilient Infrastructure}, 7(1):40--55, February 2020.

\bibitem{Citraro2020}
Salvatore Citraro and Giulio Rossetti.
\newblock Identifying and exploiting homogeneous communities in labeled networks.
\newblock {\em Applied Network Science}, 5(1), August 2020.

\bibitem{Blondel2008}
Vincent~D Blondel, Jean-Loup Guillaume, Renaud Lambiotte, and Etienne Lefebvre.
\newblock Fast unfolding of communities in large networks.
\newblock {\em Journal of Statistical Mechanics: Theory and Experiment}, 2008(10):P10008, October 2008.

\bibitem{Grootendorst2022}
Maarten Grootendorst.
\newblock Bertopic: Neural topic modeling with a class-based tf-idf procedure, 2022.

\bibitem{McInnes2018}
Leland McInnes, John Healy, Nathaniel Saul, and Lukas Gro{\ss}berger.
\newblock {UMAP}: Uniform manifold approximation and projection.
\newblock {\em Journal of Open Source Software}, 3(29):861, September 2018.

\bibitem{Bellman1966}
Richard Bellman.
\newblock Dynamic programming.
\newblock {\em Science}, 153(3731):34--37, 1966.

\bibitem{Dieng2020}
Adji~B. Dieng, Francisco J.~R. Ruiz, and David~M. Blei.
\newblock Topic modeling in embedding spaces.
\newblock {\em Transactions of the Association for Computational Linguistics}, 8:439--453, 2020.

\bibitem{Bouma2009}
Gerlof Bouma.
\newblock Normalized (pointwise) mutual information in collocation extraction.
\newblock {\em Proceedings of GSCL}, 30:31--40, 2009.

\bibitem{Grootendorst2020keybert}
Maarten Grootendorst.
\newblock Keybert: Minimal keyword extraction with bert., 2020.

\bibitem{morini2020capturing}
Virginia Morini, Laura Pollacci, and Giulio Rossetti.
\newblock Capturing political polarization of reddit submissions in the trump era.
\newblock In {\em SEBD}, pages 80--87, 2020.

\bibitem{pewResearch2017}
1615 L.~St {NW}, Suite~800 Washington, and {DC} 20036 {USA}202-419-4300 \{{\textbackslash}textbar\} Main202-857-8562 \{{\textbackslash}textbar\} Fax202-419-4372 \{{\textbackslash}textbar\}~Media Inquiries.
\newblock 1. partisan divides over political values widen.

\bibitem{redditStats}
Top websites ranking - most visited websites in june 2023.

\bibitem{Fellbaum1998}
Christiane Fellbaum, editor.
\newblock {\em WordNet: An Electronic Lexical Database}.
\newblock Language, Speech, and Communication. MIT Press, Cambridge, MA, 1998.

\bibitem{Lloyd1982}
S.~Lloyd.
\newblock Least squares quantization in pcm.
\newblock {\em IEEE Transactions on Information Theory}, 28(2):129--137, 1982.

\bibitem{Carbonell1998}
Jaime Carbonell and Jade Goldstein.
\newblock The use of {MMR}, diversity-based reranking for reordering documents and producing summaries.
\newblock In {\em Proceedings of the 21st annual international {ACM} {SIGIR} conference on Research and development in information retrieval}, page 335–336. {ACM}, August 1998.

\bibitem{Jhaver2018}
Shagun Jhaver, Larry Chan, and Amy Bruckman.
\newblock The view from the other side: The border between controversial speech and harassment on kotaku in action.
\newblock {\em First Monday}, February 2018.

\bibitem{Massanari2016}
Adrienne Massanari.
\newblock {\#}gamergate and the fappening: How reddit's algorithm, governance, and culture support toxic technocultures.
\newblock {\em New Media and Society}, 19(3):329--346, July 2016.

\bibitem{Alturayeif2023}
Nora Alturayeif, Hamzah Luqman, and Moataz Ahmed.
\newblock A systematic review of machine learning techniques for stance detection and its applications.
\newblock {\em Neural Computing and Applications}, 35(7):5113--5144, January 2023.

\end{thebibliography}
\clearpage
\appendix
\section{Topic modeling on polarized systems}
This section shows the resulting figures from the BERTopic application for the echo chambers and communities discussed in Section \ref{sec:topicModeling} of the paper.
\subsection{Gun control}
The bar plots included in Figure \ref{fig:bertopicGuncontrol} show the evolution and distribution of the five most prominent topics identified by BERTopic within the community focusing on the war in Syria. Each bar plot refers to a specific temporal span outlined in the caption.
\vspace{5em}
\begin{figure}[H]
\centering
    \subfloat[01/01/18 - 01/07/18]{
        \includegraphics[width=0.82\textwidth]{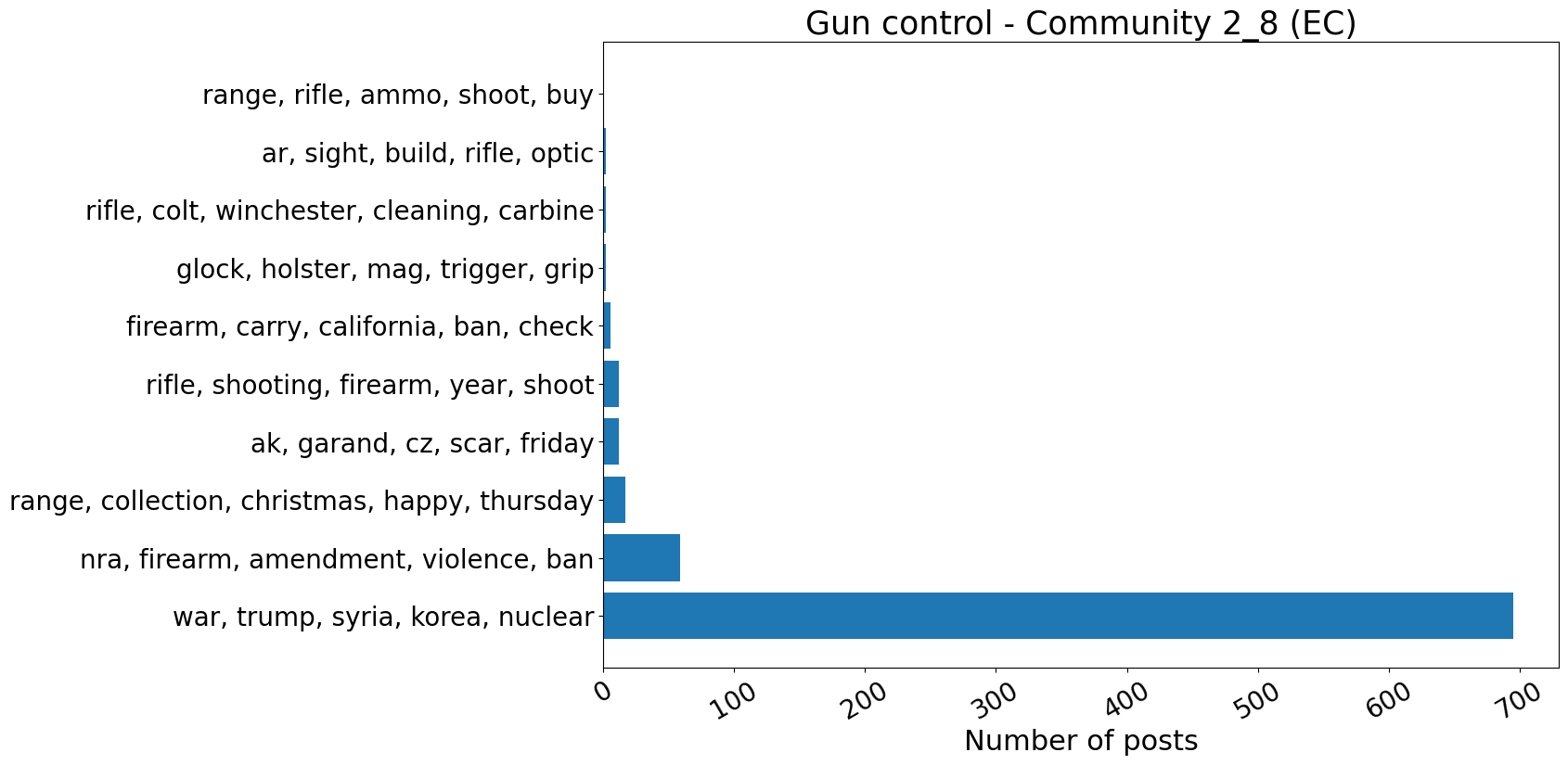}
        }
        \qquad
         \subfloat[01/07/18 - 01/01/19]{
        \includegraphics[width=0.82\textwidth]{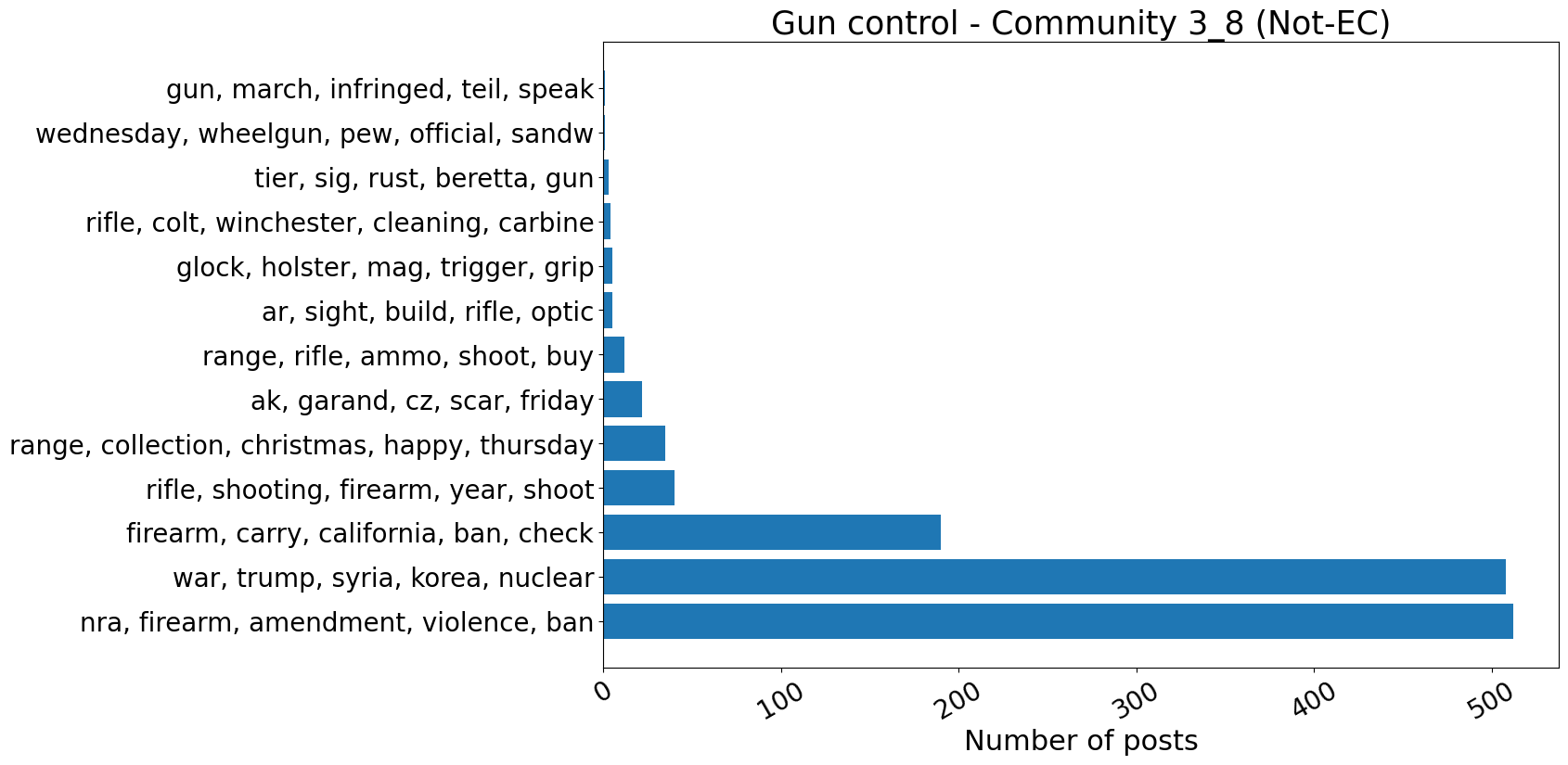}
        }
\end{figure}
\begin{figure}[H]
\centering
\ContinuedFloat
    \subfloat[01/01/19 - 01/07/19]{
    \includegraphics[width=0.82\textwidth]{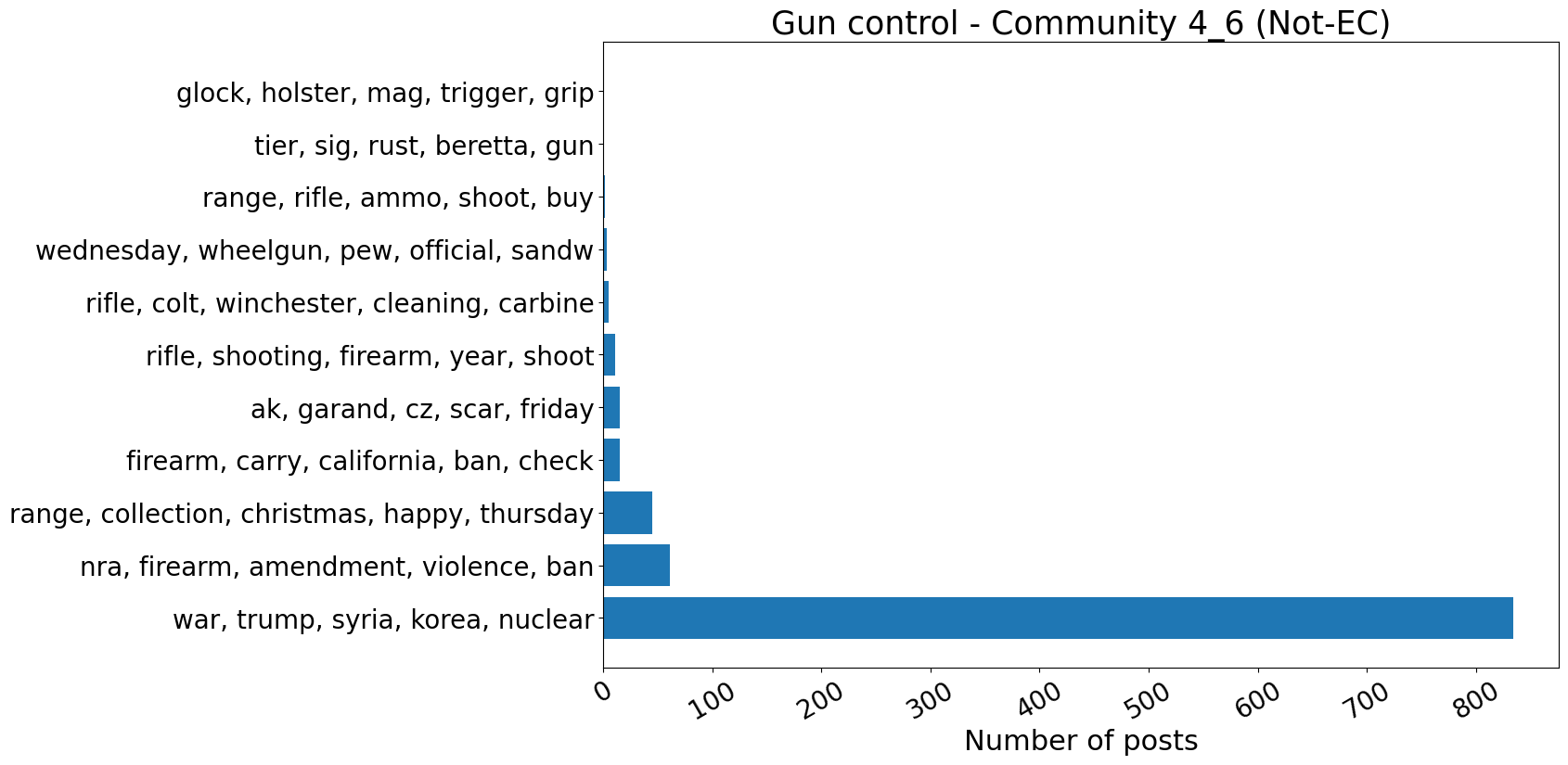}
    }
    \setcounter{figure}{3}
    \caption{For each semester, distribution of the topics in communities discussing War in Syria (Gun control)}
    \label{fig:bertopicGuncontrol}
\end{figure}

\subsection{Minorities discrimination}
The plot represents the unfolding of the discussions about Gamergate (Figure \ref{fig:bertopicMinorityGG}) and antifascism (Figure \ref{fig:MinorityBerkeley}) over the two years and a half considered by the case study, as described in Section \ref{sec:topicModeling} of the paper.
\begin{figure}[H]
\centering
\setcounter{subfigure}{0}
    \subfloat[01/17 - 07/17]{
      \includegraphics[width=0.82\textwidth]{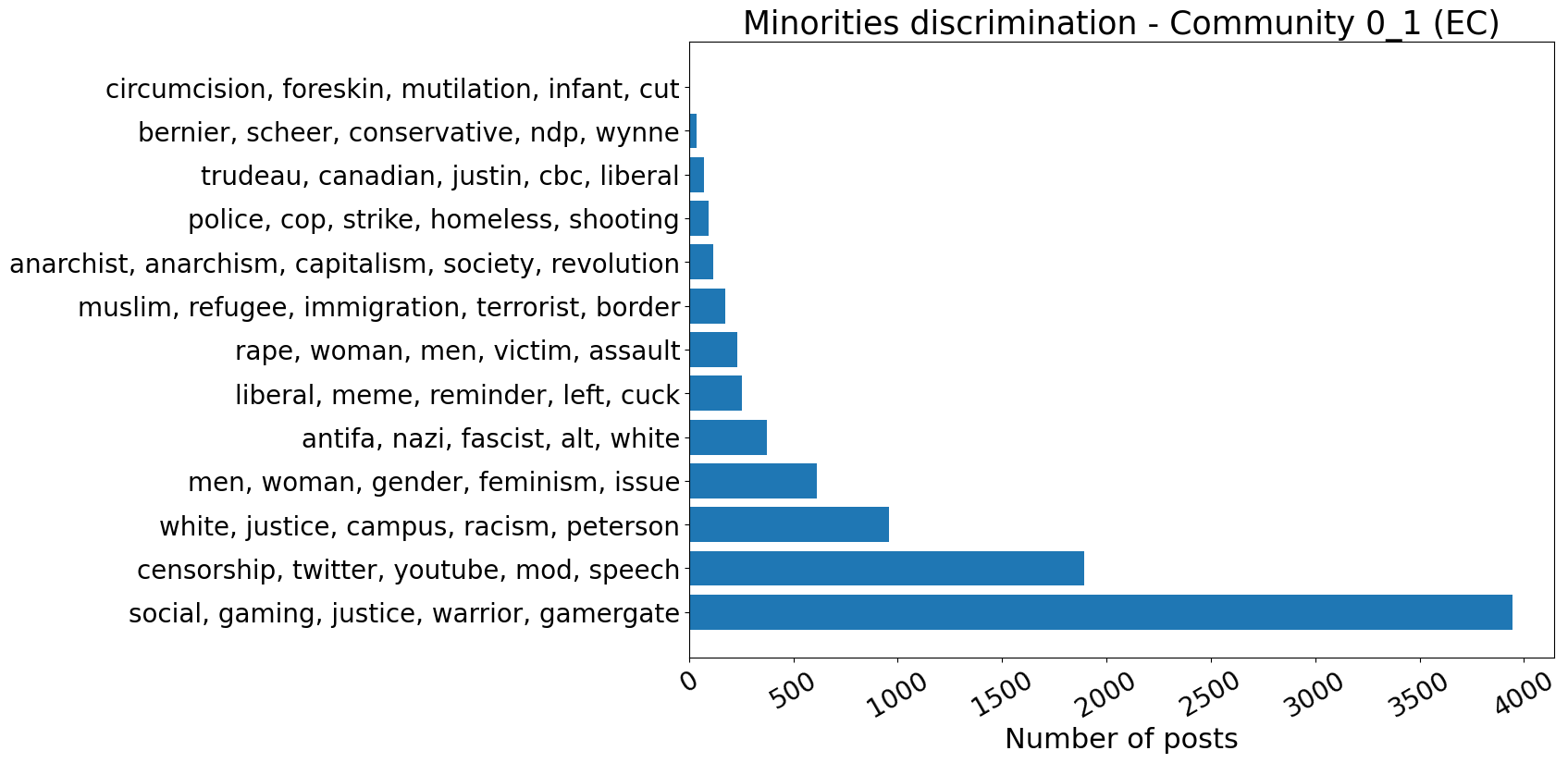}
        }
    \qquad
    \subfloat[07/17 - 01/18]{
    \includegraphics[width=0.82\textwidth]{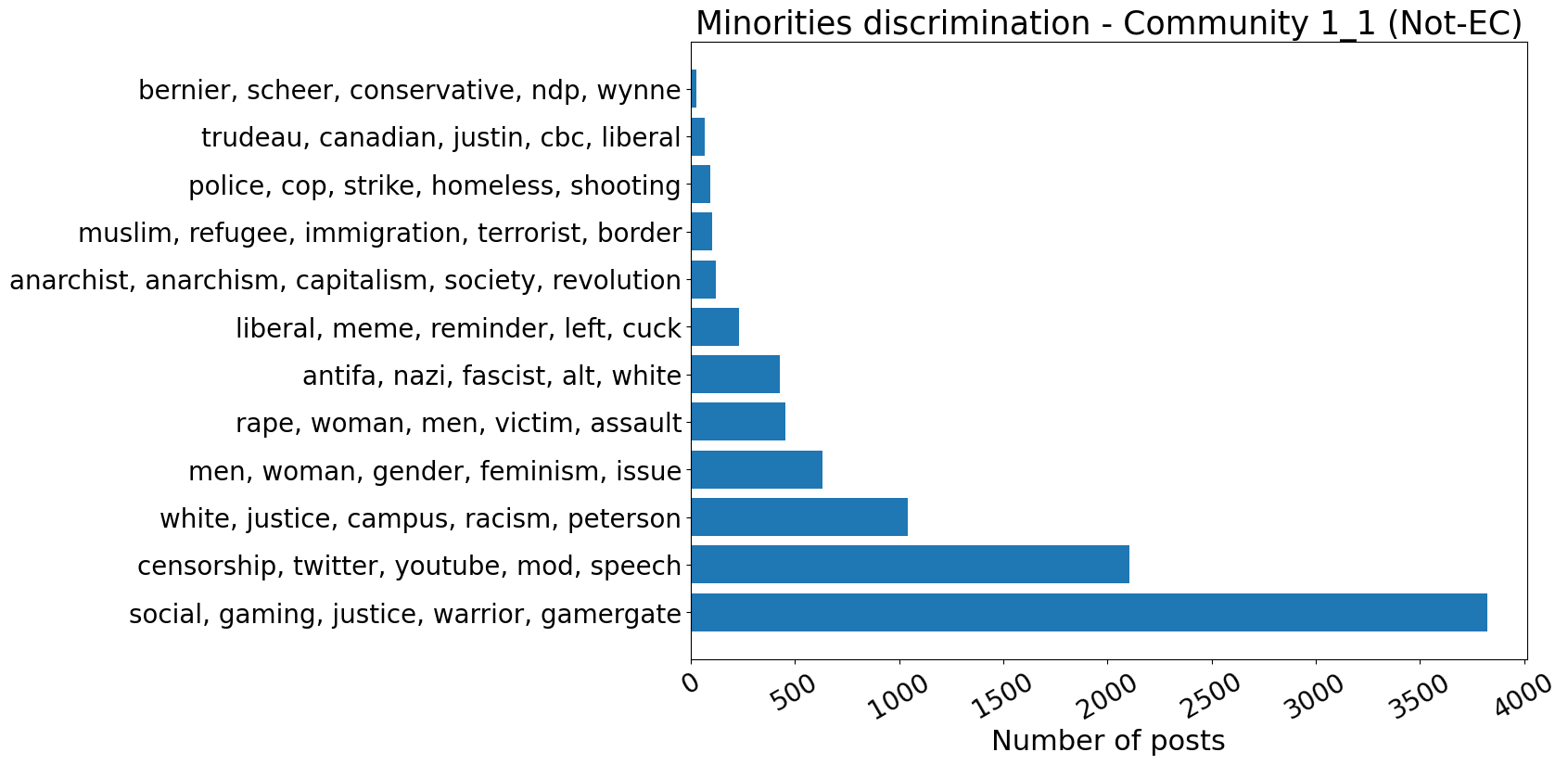}
        }
    \end{figure}
    
    \begin{figure}[H]   \ContinuedFloat
    \centering
    \subfloat[01/18 - 07/18]{
    \includegraphics[width=0.82\textwidth]{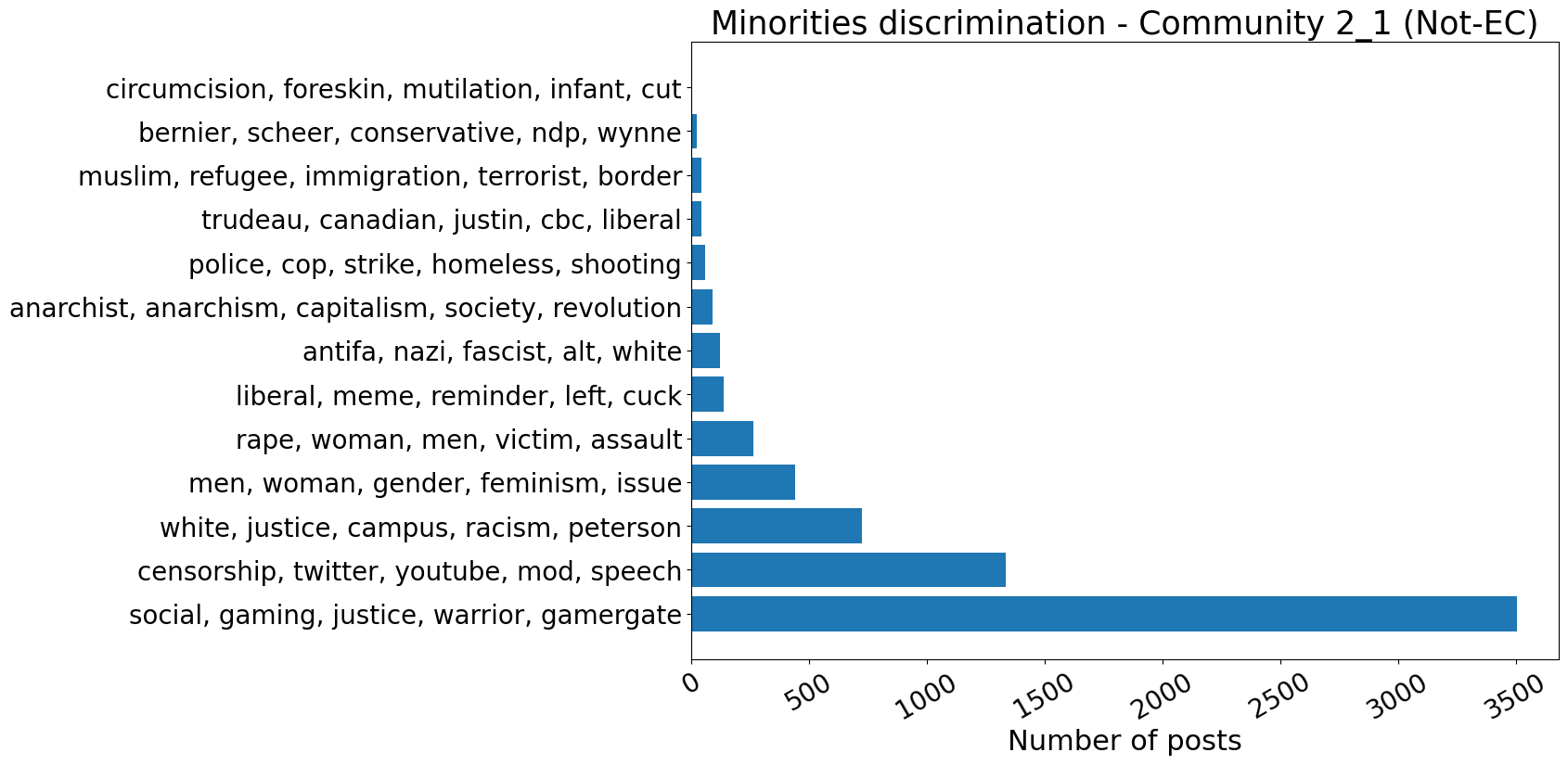}
    }
    \qquad
        \subfloat[07/18 - 01/19]{
    \includegraphics[width=0.82\textwidth]{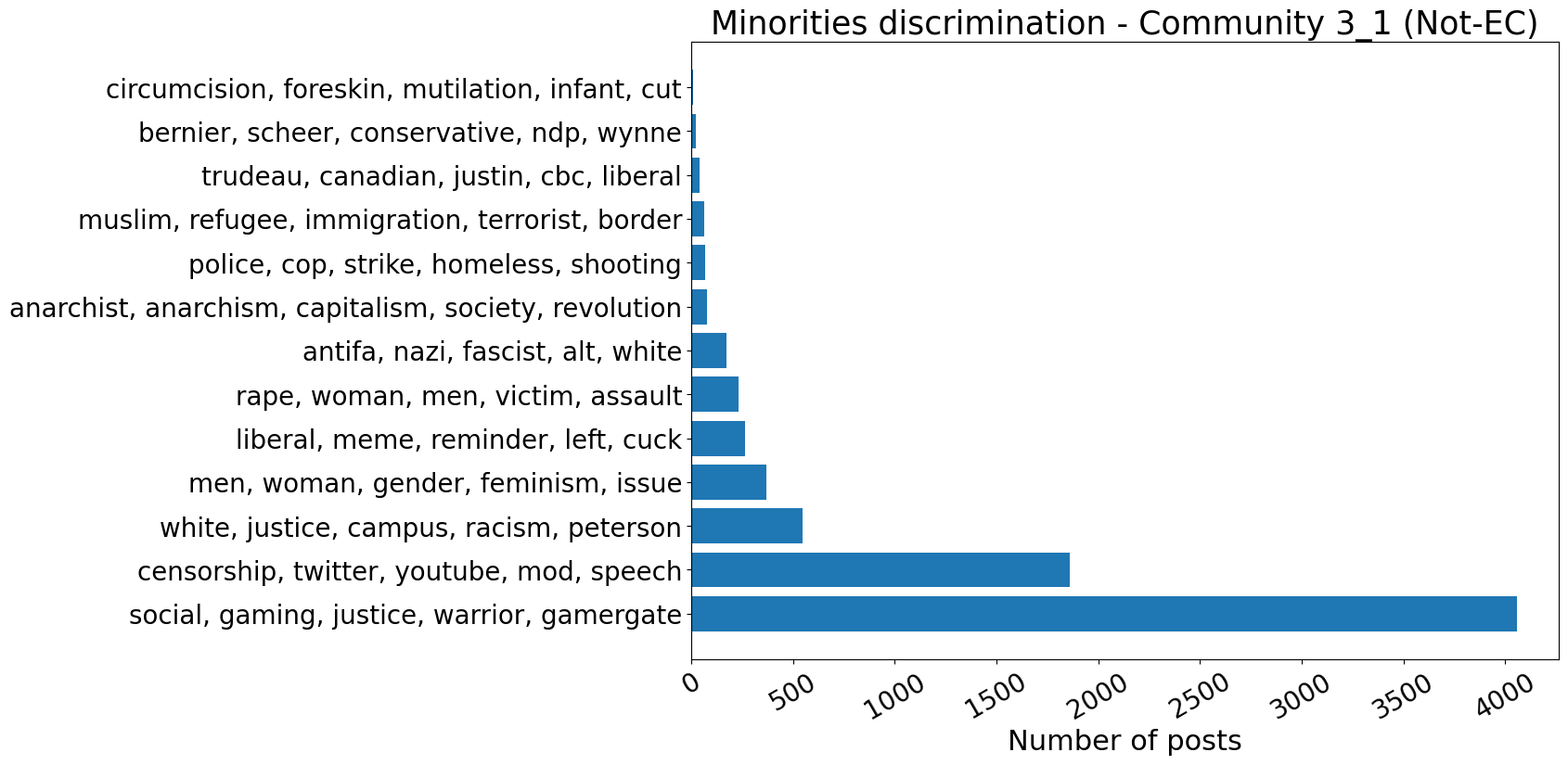}
        }
        \qquad
        \subfloat[01/19 - 07/19]{
        \includegraphics[width=0.82\textwidth]{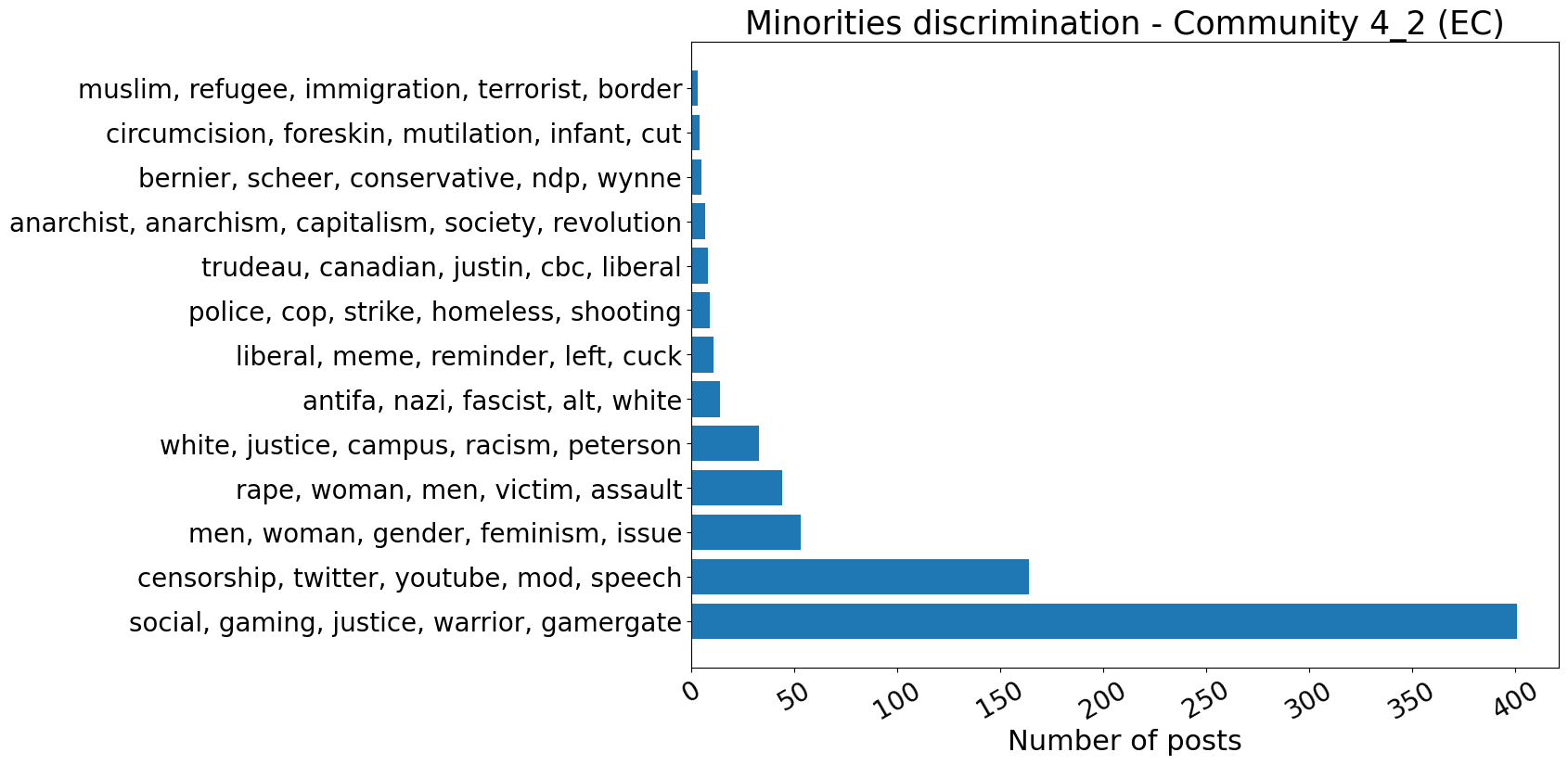}
        }
        \setcounter{figure}{4}
        \caption{For each semester, distribution of the topics in communities discussing \textit{Gamergate} (Minorities discrimination)}
        \label{fig:bertopicMinorityGG}
    \end{figure}

\begin{figure}[H]
\centering
\setcounter{subfigure}{0}
    \subfloat[01/17 - 07/17]{
    \includegraphics[width=0.82\textwidth]{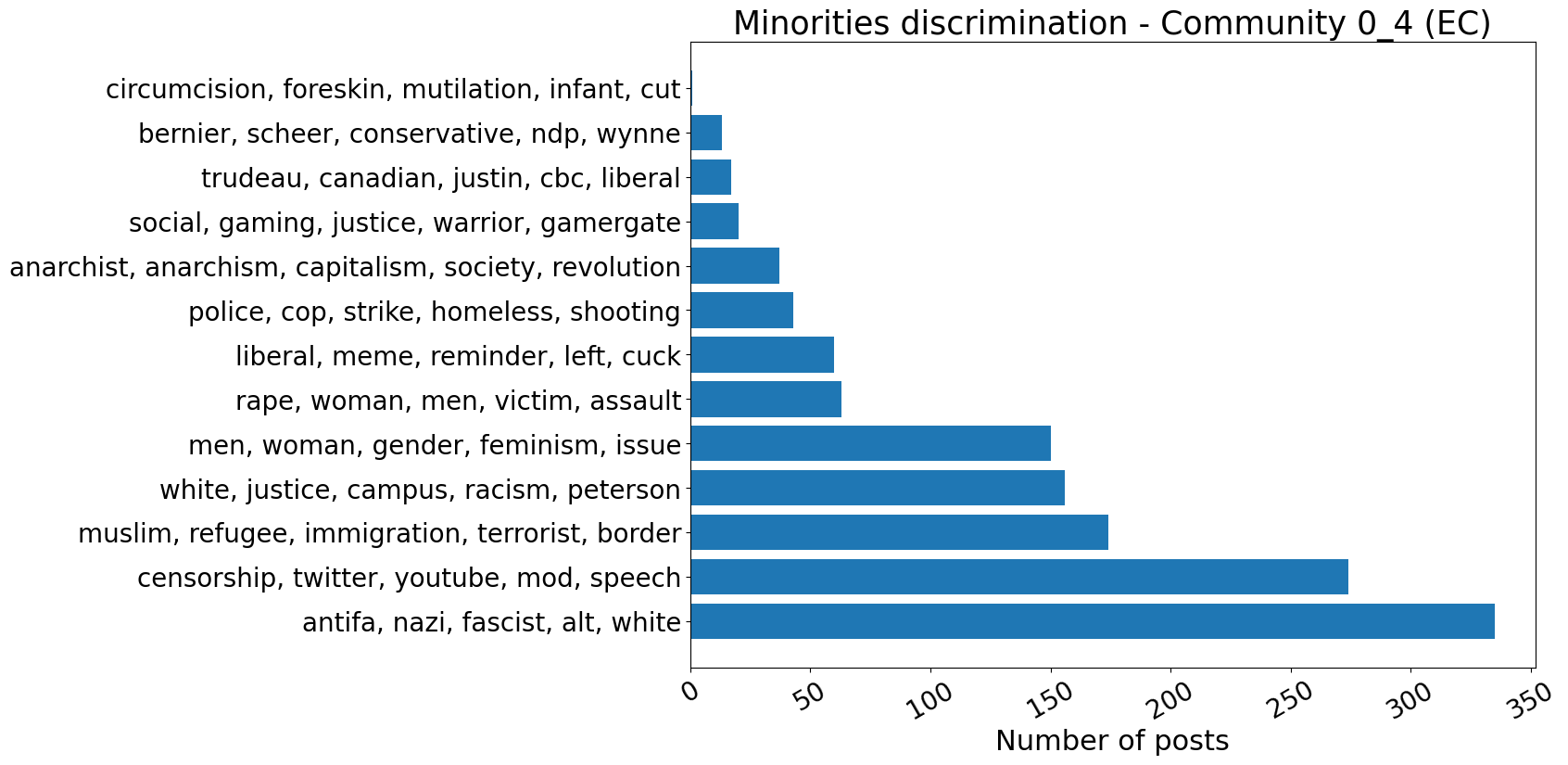}
    }
   \qquad
   \subfloat[07/17 - 01/18]{
    \includegraphics[width=0.82\textwidth]{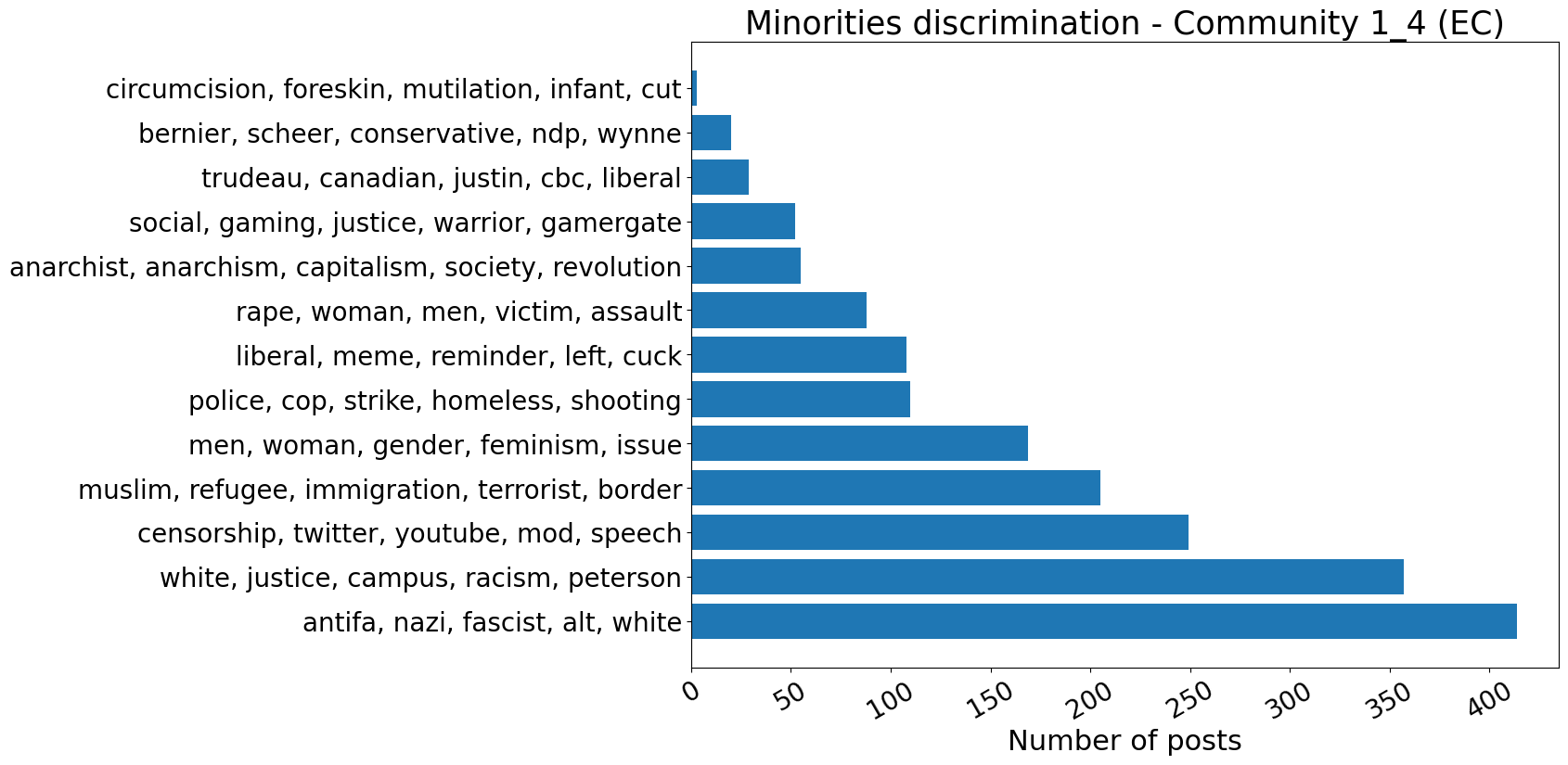}
    }
     \qquad
   \subfloat[01/18 - 07/18]{
    \includegraphics[width=0.82\textwidth]{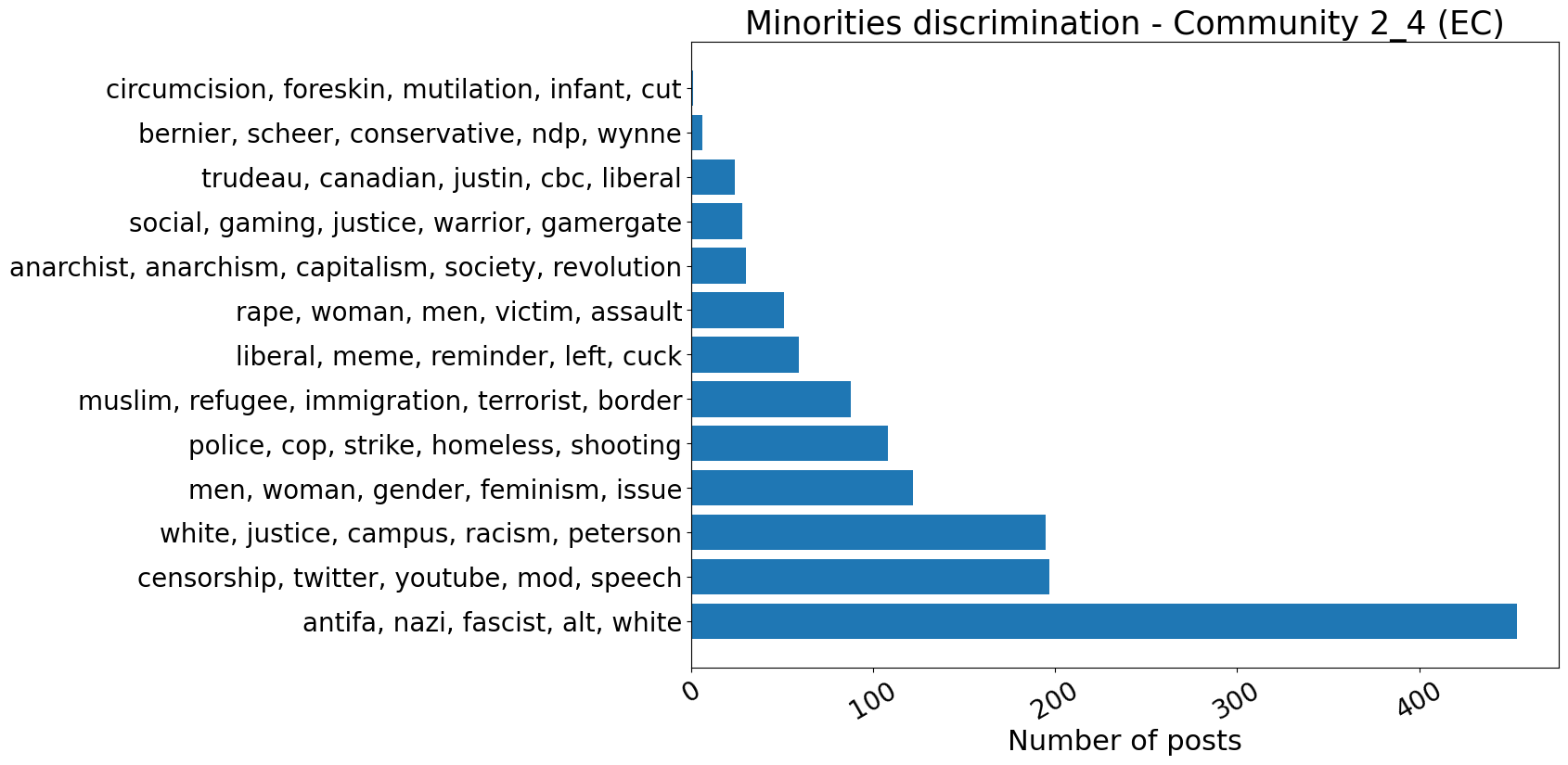}
    }
     \end{figure}
    
    \begin{figure}[H]\ContinuedFloat
    \centering
   \subfloat[07/18 - 01/19]{
    \includegraphics[width=0.82\textwidth]{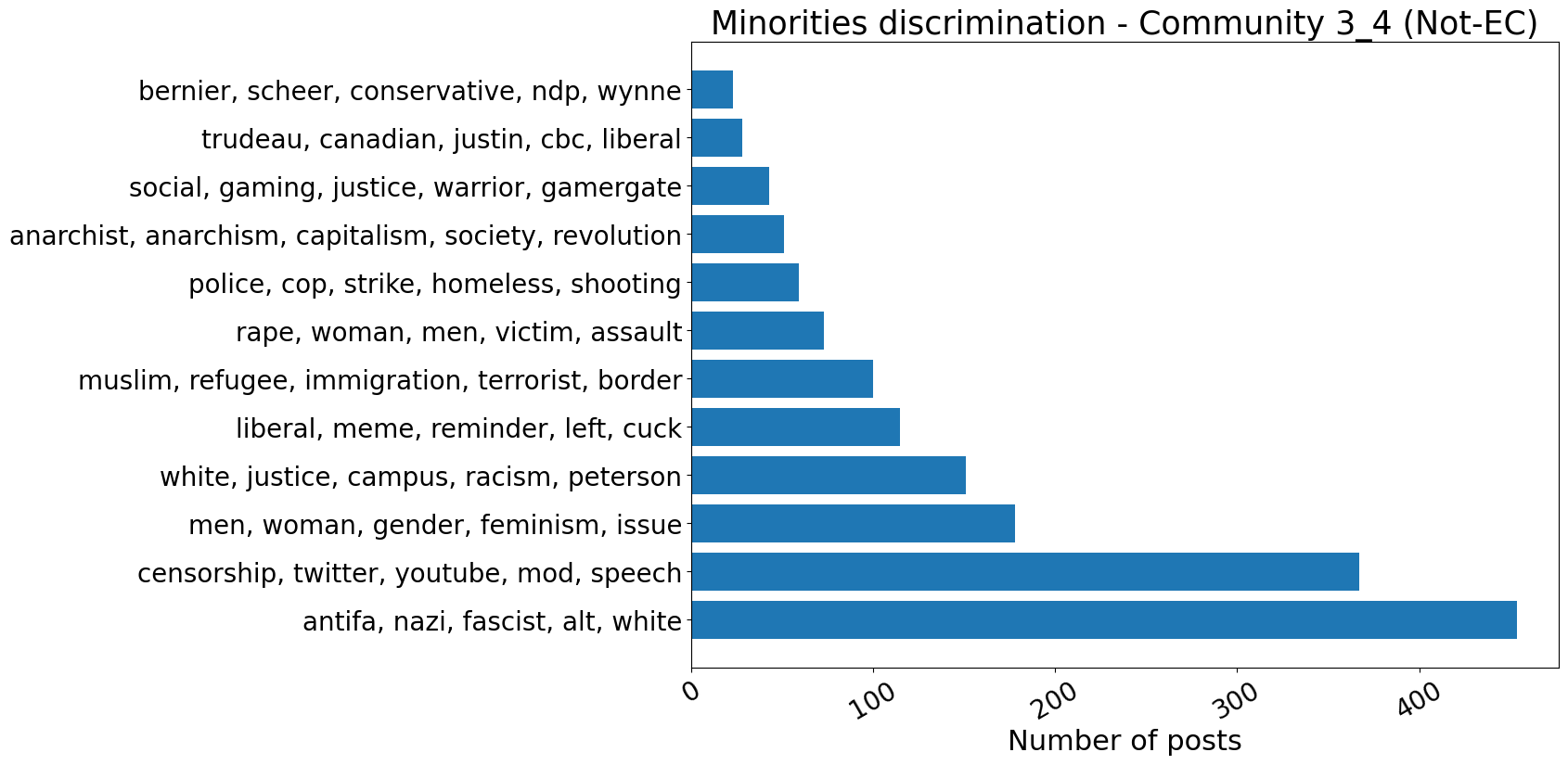}
    }
     \qquad
   \subfloat[01/19 - 07/19]{
    \includegraphics[width=0.82\textwidth]{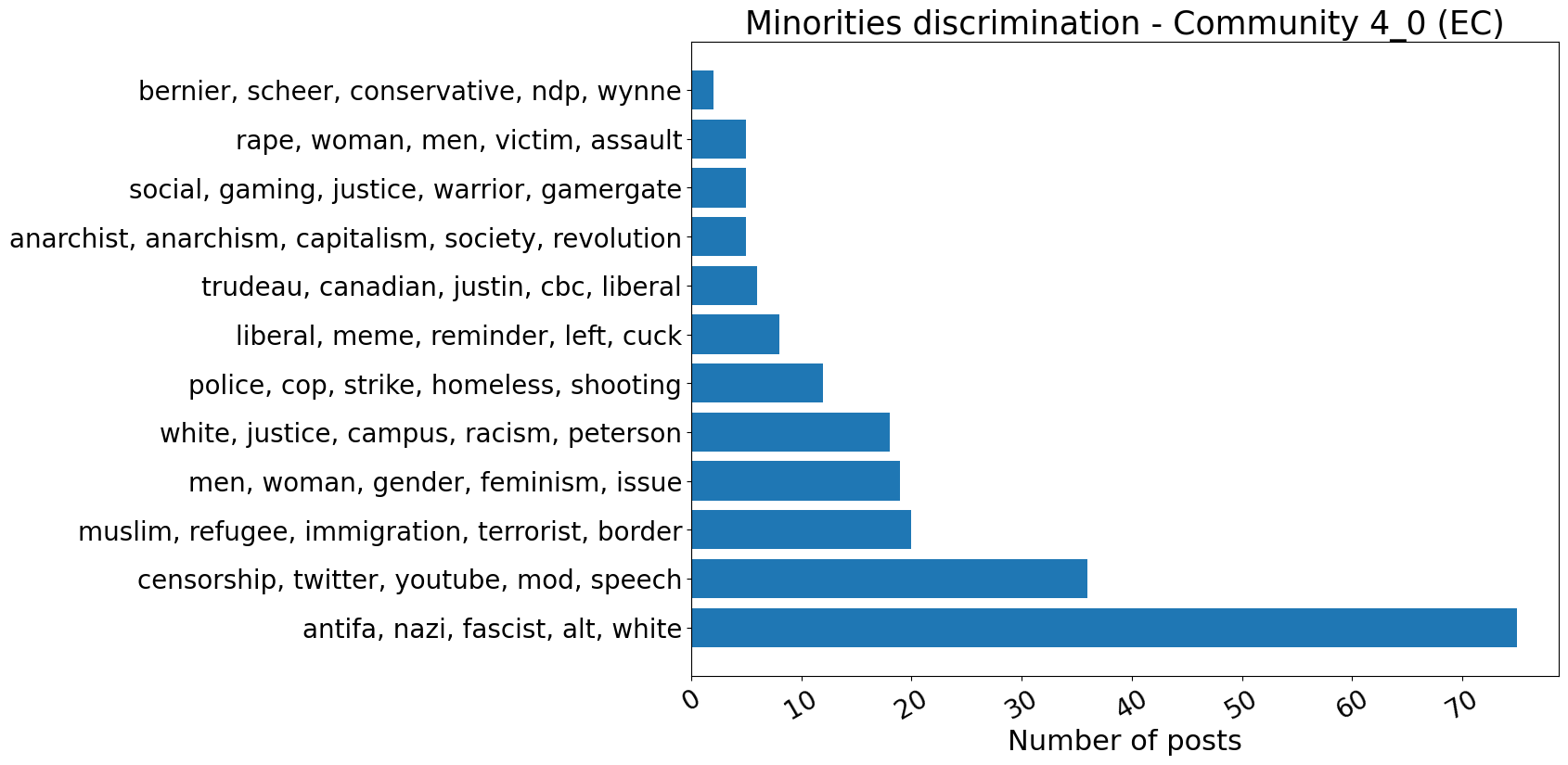}
    }
    \setcounter{figure}{5}
    \caption{For each semester, distribution of the topics in communities discussing Antifascism (Minorities discrimination)}
   \label{fig:MinorityBerkeley}
\end{figure}

\clearpage
\subsection{Politics}
Figure \ref{fig:bertopicPolitics} shows bar plots revolving around the \textit{Mueller special counsel investigation}.
\begin{figure}[H]
\centering
\setcounter{subfigure}{0}
    \subfloat[07/17 - 01/18]{
    \includegraphics[width=0.82\textwidth]{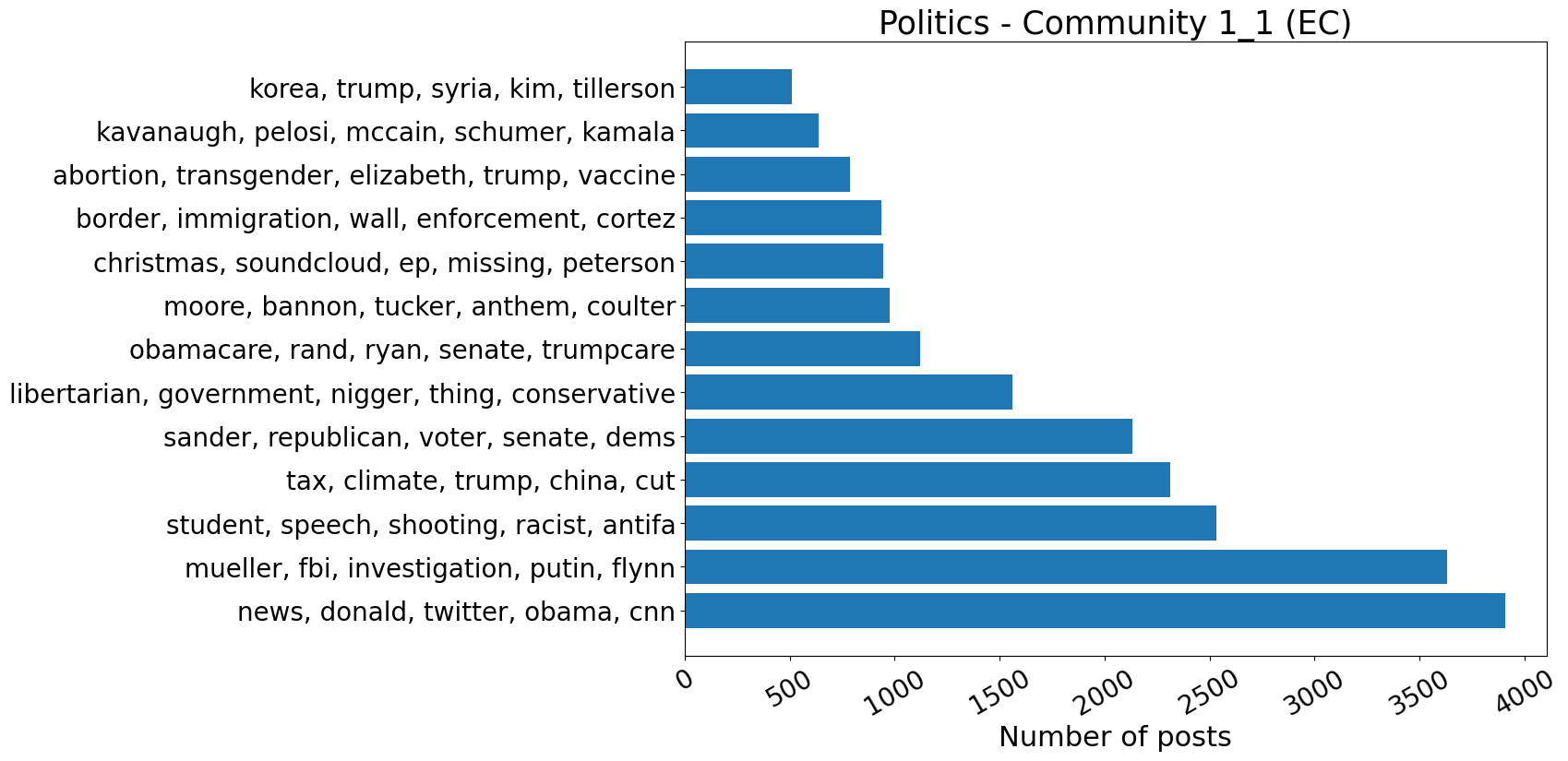}
    }
   \qquad
   \subfloat[01/18 - 07/18]{
    \includegraphics[width=0.82\textwidth]{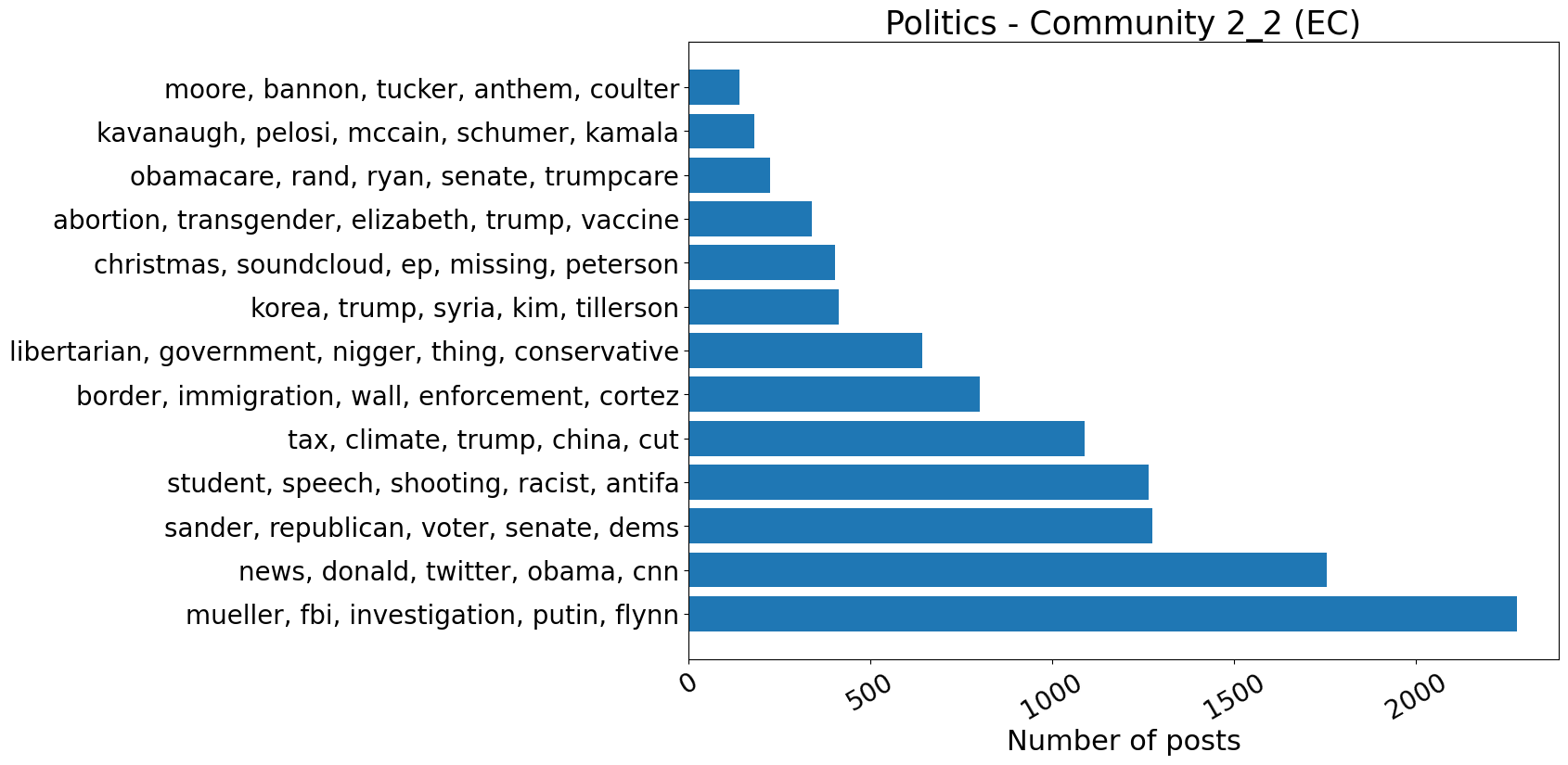}
    }
    \qquad
    \subfloat[07/18 - 01/19]{
    \includegraphics[width=0.82\textwidth]{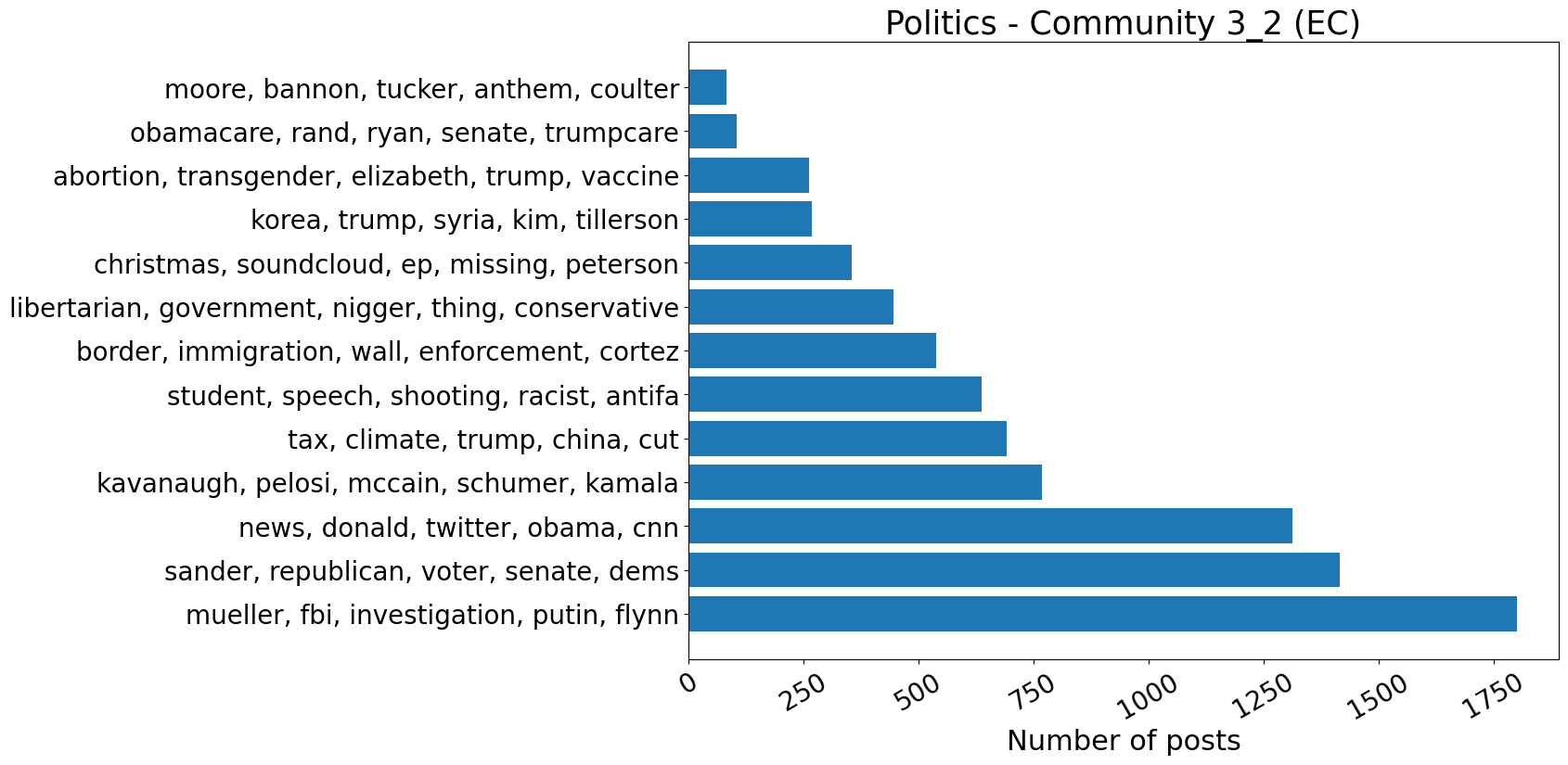}
    }
\end{figure}
\begin{figure}[H]\ContinuedFloat
    \centering
    \qquad
   \subfloat[01/19 - 07/19]{
    \includegraphics[width=0.82\textwidth]{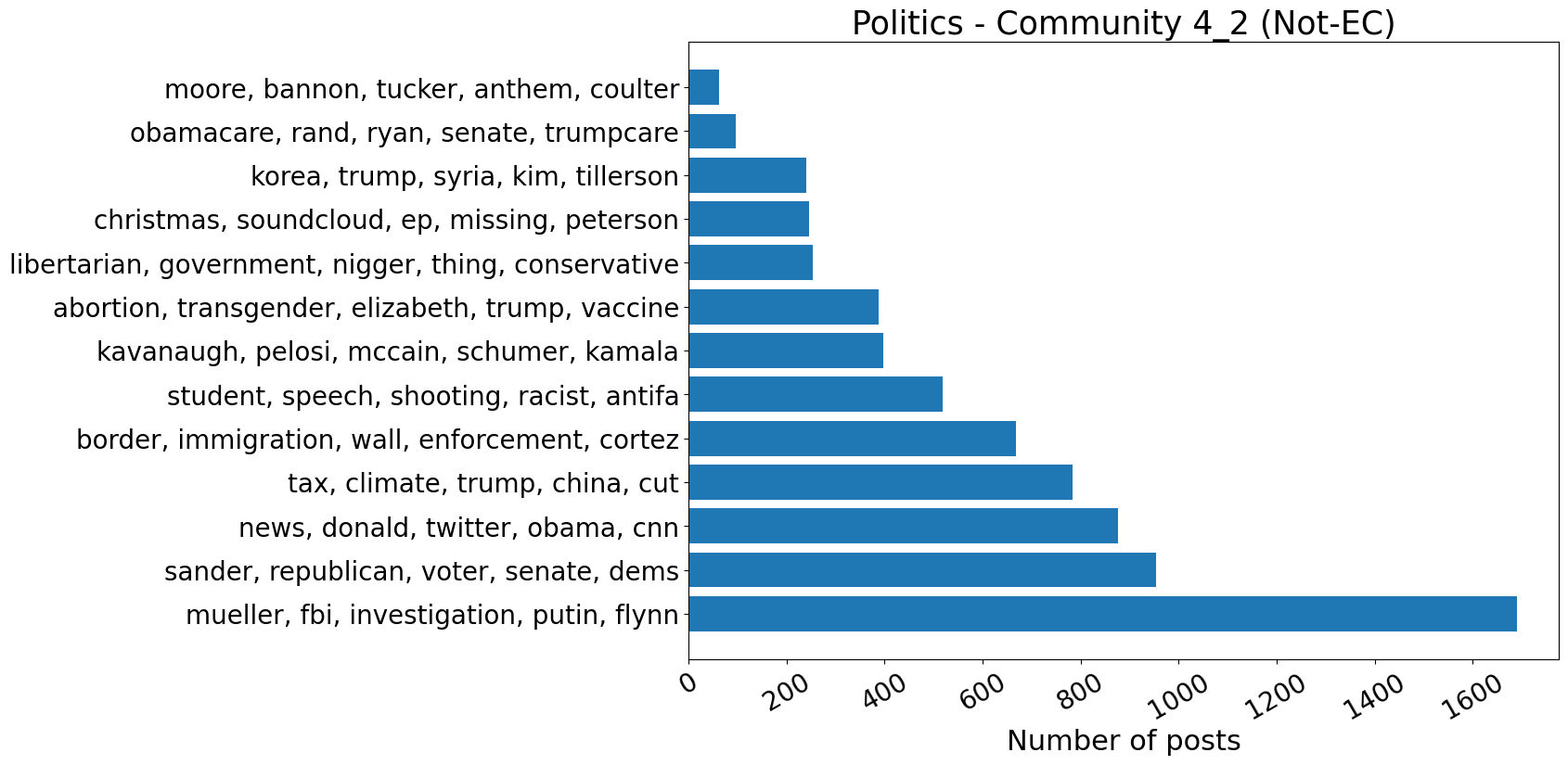}
    }
    \setcounter{figure}{6}
    \caption{For each semester, extracted topics in communities discussing about \textit{Mueller special counsel investigation}}
    \label{fig:bertopicPolitics}
    
\end{figure}

\end{document}